\documentclass[twocolumn,twocolappendix,tighten]{openjournal}
\usepackage{graphicx}
\usepackage{xspace}
\usepackage{bm}
\usepackage{bbm}
\usepackage{amsmath}
\usepackage{tikz}
\RequirePackage[colorlinks=true,linkcolor=blue,citecolor=blue,urlcolor=blue]{hyperref}

\begin{document}

\title[]{Turbulence in the multi-phase circumgalactic medium: \\bridging TNG50 and idealized kpc-scale simulations}
\author{Jonas Biba$^{1,\star}$ and Dylan Nelson$^{1,2,\dagger}$}

\affiliation{$^{1}$ Universit\"{a}t Heidelberg, Institut f\"{u}r Theoretische Astrophysik, Zentrum f\"{u}r Astronomie, AU-Str. 2, 69120 Heidelberg, Germany}
\affiliation{$^{2}$ Universität Heidelberg, Interdisziplinäres Zentrum für Wissenschaftliches Rechnen, INF 205, 69120 Heidelberg, Germany}
\thanks{$^\star$ E-mail: \href{mailto:jonas.biba@stud.uni-heidelberg.de}{jonas.biba@stud.uni-heidelberg.de}}
\thanks{$\dagger$ E-mail: \href{mailto:dnelson@uni-heidelberg.de}{dnelson@uni-heidelberg.de}}

\begin{abstract}
The circumgalactic medium (CGM) is the extended multi-phase gaseous halo surrounding a galaxy. To study how its dynamics are shaped by the interplay of radiative cooling and turbulence, we first turn to the TNG50 cosmological magnetohydrodynamical simulation. We analyze the turbulent properties of the CGM of Milky Way-like galaxies at $z=0$, using a multi-scale filtering technique. We find that the CGM hosts predominantly subsonic flows with Mach number $\mathcal{M} \sim 0.1 - 0.7$ that are roughly balanced between compressive and solenoidal components with $b = \mathcal{M}_{\rm comp}/\mathcal{M}_{\rm tot} \sim 0.3 - 0.6$. We then use these physically motivated properties of turbulent CGM media to set the initial conditions and driving parameters for idealized, turbulent box simulations. These represent $\sim \rm{1\, kpc}^3$ volumes of the CGM, but reach resolutions higher than available in cosmological volumes such as TNG50. To do so, we implement a turbulent driving algorithm into the AREPO code based on random but smoothly time-fluctuating acceleration fields, and then run a suite of turbulent boxes across the parameter space, including a model for metal-line radiative cooling. Using these simulations, we study how turbulence seeds density fluctuations that trigger thermal instability and promotes mixing between hot and cold phases, depending on the relative timescales of cooling and mixing. The abundance and evolution of cold $\sim 10^4$\,K gas depends strongly on the strength of turbulence and the efficiency of cooling.
\end{abstract}



\section{Introduction}

The circumgalactic medium (CGM) is the gaseous halo surrounding galaxies. It plays a central role in galaxy formation and evolution, mediating important physical processes \citep{fauchergiguere2023}. Gas accretion from both the CGM and the intergalactic medium (IGM) supplies the fuel for star formation \citep{Tumlinson_2017}. At the same time, feedback from active galactic nuclei (AGN) and stellar-driven outflows, including those due to supernovae, deposit mass and energy back into the CGM. A complete picture of galaxy formation therefore necessitates a rigorous understanding of the physical properties and physical processes of the CGM.

Observations reveal that the CGM is not only multi-phase \citep{DONAHUE20221}, but also turbulent \citep{10.1093/mnras/stac3193}, suggesting gas flows with high Reynolds numbers and significant velocity fluctuations \citep{Elmegreen_2004,Burkhart_2021,shah-turbulentcgmmixing}. The two are intricately linked, and turbulence is a small-scale process that regulates both hot and cold phases and their interactions in the CGM. Overall, the presence of turbulence modulates the dynamic and thermal properties of the CGM. Heating through turbulent dissipation counteracts radiative cooling \citep{Zhuravleva2014}, while turbulent motions can also seed density fluctuations that trigger the formation of cold gas through thermal instability \citep[TI;][]{Mohapatra_2021}. Mixing pre-existing cold gas with hot gas in turbulence interface layers can produce thermally unstable intermediate temperature gas that grows cold gas clouds through condensation \citep{Fielding2020}. 

However, any process that suppresses unstable regions will counteract the formation of a multi-phase medium, where turbulent mixing is one such example. The interaction of radiative cooling and turbulence in the CGM therefore sets key details about the composition, morphology, and evolution of the CGM.

Numerical simulations that probe small-scale gas physics including turbulence come in two fundamental flavors. First, cosmological simulations model galaxies and their gas starting from initial conditions consistent with our $\Lambda$CDM concordance model. Over the past two decades, these have become powerful tools to study galaxy populations as a whole, self-consistently capturing the coupled evolution of dark matter, gas, stars, supermassive black holes within the diversity of cosmic environments \citep{Vogelsberger2020}, and the impact of galactic processes such as feedback physics on the realistic CGM \citep{Rey2025}. Simultaneously, high-resolution simulations can follow small-scale details and gas physics, enabling them to begin to resolve the important role of turbulence \citep{lochhaas23,goldner25}. For example, the large-volume TNG50 simulation resolves an abundant formation of cold clouds in massive hot halos due to triggered TI from strong density perturbations \citep{Nelson-coldgas}, while the GIBLE simulations reveal how magnetic field draping effects the evolution of CGM clouds \citep{Ramesh2024,Ramesh2026}. 

On the other hand, idealized simulations offer more controlled environments. These numerical experiments rely on carefully chosen initial conditions, and are sufficiently flexible (and computationally inexpensive) to explore a wide range of physics and physical parameters. Idealized simulations have shown how turbulence regulates the energy, density, and momentum structure of gas across a variety of environments, from the star-forming ISM \citep{MacLow04}, to the CGM \citep{fauchergiguere2023}, to the hot intracluster medium (ICM) of galaxy clusters (\citealt{Iapichino08}, \textcolor{blue}{Saha et al. in prep}). In particular, the role of turbulence in the star-forming ISM, and its role in regulating the process of star formation, is well studied with simulations of driven turbulence \citep{stone98}. These works have shown how turbulence regulates the gas density distribution, velocity statistics, and energy dissipation rate, setting the stage for models of cloud-scale star formation rate \citep{vasquez03,Federrath_2012}. At the other extreme, turbulence in virialized, hot plasma of the ICM is understood to play a key role in the interplay of heating versus radiative cooling and large-scale cooling flows \citep{gaspari13,Mohapatra_2020,lehle25}, the meditation of feedback energy from AGN \citep{li14}, modifications to the impact of thermal conduction \citep{Ruszkowski10}, and local thermal instability \citep{sharma10}.

The circumgalactic medium is a complex middle ground, both in terms of the halo mass scale, and the relevant physics \citep{voit18}. Even though the injection mechanism of turbulence in galaxies might be reasonably resolved in cosmological simulations, they cannot fully resolve the large dynamic range in e.g. gas density, temperature, and cooling time that is present \citep{suresh19}. This makes idealized simulations essential. Local CGM calculations are frequently used to study the evolution of pre-existing cool clouds within hot media \citep{armillotta17,sparre20} and the interface layers between phases \citep{Ji__2019,Fielding2020,mandelker20}. These geometries and setups have key differences with respect to so-called `driven turbulent boxes', including in the interpretation and physical evolution of fundamental statistics such as the temperature distribution \citep{chen26}. Simulations of (driven) turbulent media in CGM-like conditions are a less explored regime. These have focused on the formation, evolution, and survival of cool gas \citep{gronke22}, as well as the observable metal ion signatures \citep{buie18,buie20,buie20b,hu25}.

In this work, we aim to combine these two approaches. First, we analyze the CGM properties in the TNG50 cosmological galaxy formation simulation, and particularly focus on its turbulent motion. Large scale simulations such as TNG50 are useful since the state of the CGM emerges as a consequence of the global cosmological evolution of galaxies. For example, the history of mergers, presence of satellite galaxies, and feedback from supernovae and SMBHs can all impact the physical state of the CGM. We then use several representative states as initial conditions for idealized, small-scale turbulent simulations that can then reproduce realistic conditions in the CGM. To do so, we drive turbulent motions statistically consistent with those seen in the cosmological simulations. Including radiative cooling, we then investigate the properties and phase structure of the multi-phase, turbulent CGM.

In the Kolmogorov picture of turbulence, the dynamics at scales below the energy injection scale are universal. This makes the analysis of small-scale structure of turbulence broadly applicable, independent of the specific physical mechanism driving the turbulence. However, as the Reynolds number can be as high as \(\rm{Re} \sim 10^7\) \citep{Elmegreen_2004}, the range of dynamically relevant scales is difficult to represent in direct numerical simulations. This range grows with the Reynolds number as $L/\eta \sim \rm{Re}^{3/4}$, where $L$ is the injection scale and $\eta$ is the length-scale of physical viscous dissipation. Instead of explicitly resolving the physical dissipation scale, we allow numerical viscosity take the place of the physical viscosity, solving the Euler equations instead of the Navier-Stokes equations \citep{SCHMIDT2006353}. This is a widely used approach and provides a useful approximation, as explicitly including viscosity would require a more detailed understanding of the physical mechanisms governing viscous dissipation in the CGM, while some degree of numerical dissipation is unavoidable in finite volume hydrodynamic schemes as used in this work. While numerical dissipation is not physical vicious dissipation, at sufficiently high resolution there exists a range of spatial scales analogous to the inertial range, in which the turbulent motions are independent of the dissipation processes at the grid scale.

The rest of this paper is organized as follows. Section \ref{sec_methods} introduces our methodology, including how we analyze the properties of turbulent media, the cosmological simulations we study, and the idealized turbulent box simulations that we run. In Section \ref{sec_results1} we present our results on the analysis of turbulence in the CGM of Milky Way-like galaxies from TNG50. Section \ref{sec_results2} considers idealized turbulence simulations with properties motivated by these results, while Section \ref{sec_results3} adds radiative cooling and studies the properties of cooling and cool gas in the idealized turbulent CGM. We discuss limitations and future directions in Section \ref{sec_discussion}, while Section \ref{sec_conclusions} summarizes our main findings. 


\section{Methods} \label{sec_methods}

\begin{figure*}
    \centering
    \includegraphics[width=0.99\textwidth]{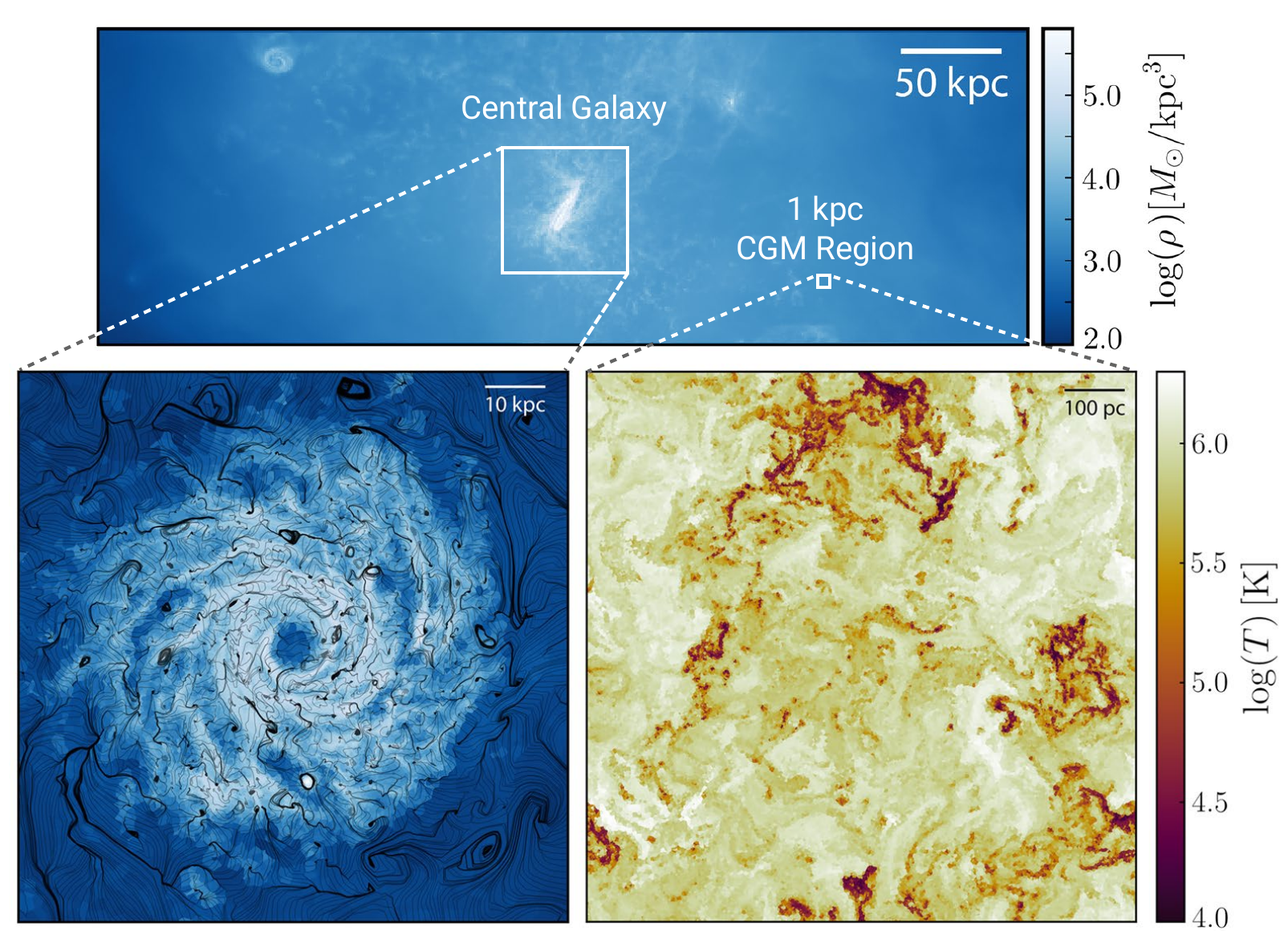}
    \caption{Schematic overview of our simulations and analyses. We start with the Milky Way-like galaxies at $z=0$ from the TNG50 cosmological simulation (top panel), shown here in gas density projection. We analyze the physical properties and characteristics of turbulence in the circumgalactic medium, in order to provide initial conditions for our idealized simulations of small-scale driven turbulence (lower right panel; color shows temperature fluctuations in the stationary phase). In order to estimate the turbulent Mach number $\mathcal{M}$ and compressive ratio parameter $b$ in TNG50, we decompose the gas velocity field into its bulk and turbulent components (lower left panel; gas density slice through the disk plane, with black streamlines tracing the turbulent velocity component).}
    \label{fig:overview}
\end{figure*}

\subsection{Milky Way-like galaxies and TNG50}

The TNG50 simulation is part of the IllustrisTNG suite of large volume, cosmological, magnetohydrodynamical simulations \citep{Pillepich_2019,Nelson_2019}. TNG simulates the coevolution of dark matter, gas, stars and black holes including gas physics and a comprehensive model for galaxy formation \citep{weinberger17,pillepich18}. Gas dynamics follow the equations of ideal magnetohydrodynamics, including self-gravity and external influences such as dark matter gravity and primordial and metal-line cooling, using the \texttt{Arepo} code. The TNG model includes key astrophysics: star formation, stellar evolution and feedback, and supermassive black hole (SMBH) formation and feedback. 

The TNG50 simulation has a unique combination of large volume and high resolution that allows for analysis of small scale CGM phenomena in a global context \citep[e.g.][]{Nelson-coldgas,ramesh23}. The simulation starts from cosmologically motivated initial conditions at $z=127$ and runs until the present day, $z=0$. The simulation volume (50 Mpc) contains \(\sim\) 20,000 galaxies at redshift zero, where the median spacial resolution of the star-forming ISM gas is \(\sim\) 100-140 pc. 

From TNG50 we analyze a selection of Milky Way (MW) and Andromeda (M31) like galaxies, hereafter referred to as the MWM31 catalog \citep{2024MNRAS.535.1721P}. Figure \ref{fig:overview} shows an example of the gas structure in the CGM of one of our TNG50 galaxies (top panel), as well as the turbulent velocities present in the galactic disk (lower left). The selection of the galaxies is based on the following criteria evaluated at $z=0$, yielding a collection of 198 galaxies:

\begin{itemize}
    \item[i)] \textit{Stellar mass}: the galaxy stellar mass is in the following range: \(M_*\left(<30\mathrm{kpc}\right) = 10^{10.5 - 11.2}M_\odot\)
    \item[ii)] \textit{Stellar morphology}: the galaxies needs to have a disc-like stellar distribution with spiral arms.
    \item[iii)] \textit{Environment} No other galaxy with stellar mass \(\leq 10^{10.5} M_\odot\) is within 500 kpc distance.
\end{itemize}

The sample contains galaxies with similar stellar properties and environments, and for each a cutout containing all gas, stars, and other matter in a volume of side length \(L_\mathrm{cutout} \approx 600\,\mathrm{kpc}\) around the galaxy is available as part of the TNG public data release \citep{nelson2021}. 

\subsection{Idealized Simulations of Driven Turbulence}

We use the \texttt{Arepo} code \citep[][the public version]{Springel_2010} to model the equations of ideal (magneto)hydrodynamics. Space is discretized into finite volumes for the conserved fluid quantities (mass, momentum and energy). These volumes are the cells of the Voronoi tesselation of a discrete set of cell-generating points. Hydrodynamical fluxes between cells are computed with a second order accurate reconstruction of the primitive variables, and an exact Riemann solver. In particular, we solve the Euler equations of ideal hydrodynamics, given by the compact form:
\begin{align}
    \frac{\partial\bm{U}}{\partial t} + \bm{\nabla} \cdot \bm{F}\left(\bm{U}\right) = \bm{S}.
\end{align}
for the state vector $\bm{U} = (\rho, \rho\bm{v}, \rho e)$ and corresponding flux $\bm{F}$ and source term $\bm{S}$,
\begin{align}
\bm{F}\left(\bm{U}\right) = \begin{pmatrix}
    \rho \bm{v} \\
    \rho \bm{v} \bm{v}^t + P \\
    \left(\rho e + P\right)\bm{v} \\
\end{pmatrix}, \quad \bm{S}\left(\bm{U}\right) = \begin{pmatrix}
    0 \\
    \rho \bm{f} \\
    \rho \bm{f} \cdot \bm{v} + \Lambda_{\rm net} \\
\end{pmatrix}
\end{align}
where $\rho$ is the mass density, $\bm{v}$ is the velocity field, $P$ is the pressure, and $e = u + \bm{v}^2/2$ is the total energy per unit mass. In addition, $\rho \bm{f}$ is the turbulent driving force, and $\Lambda_\mathrm{net}(T)$ is the cooling function, both introduced below. The thermal energy per unit mass $u$ is related to the pressure and density via an equation of state: \(P = \left(\gamma - 1\right) \rho u\) (in the case of an ideal gas), that closes the above system of equations. In our simulations, we set $\gamma = 5/3$ exclusively. For a given cell in our simulation, we calculate the adiabatic speed of sound as $c_s = \sqrt{\gamma P/\rho}$.

To keep second order accuracy in time and space, the gradients of the primitive variables of the cells are estimated based on the neighboring cells and a hybrid of the MUSCL-Hancock and a Runge-Kutta time integration scheme \citep{Pakmor_2015}. Accelerations such as an external gravity field are treated as separate source terms and handled in an operator split manner \citep[see][]{Weinberger_2020}. This means that an acceleration is first applied to cells for half a time-step, updating velocity and energy. Then, the hydrodynamics equations are evolved for a full time-step in their homogeneous form, followed by another half time-step for the external source terms. This is effectively a Leapfrog-scheme in a Kick-Drift-Kick manner, ensuring energy conservation. 

We always run volumes with 1\,kpc side-length, and adopt periodic boundary conditions. Figure \ref{fig:overview} shows an example of the temperature structure of one of our idealized turbulent box simulations (lower right).

\subsection{Cooling}

Radiative cooling of the gas is treated as a sink term in the energy conservation equation and updated at the end of every time-step. The public version of the \texttt{Arepo} code includes a method for computing the ionization state of a gas cell with a given internal energy, based on a primordial cooling network. This method calculates the balance between collisional ionization and recombination for a gas composed solely of hydrogen and helium, yielding the hydrogen number density and temperature at a specified internal energy. Since the cooling rate depends strongly on the internal energy and density, the update employs a first-order implicit integration of the cooling term. 

However, since the radiative cooling also depends on the metallicity in the presence of metals, the calculation assuming a primordial composition is only an approximation. To include metal-line cooling, we use precomputed values of the net cooling function \(\Lambda_\mathrm{net}\), tabulated as a function of temperature, at fixed metallicity. In particular, \citet{2009A&A...508..751S} use the \texttt{SPEX} package to calculate the ionization state and cooling rates of the above processes for a plasma with solar metallicity and solar abundance of 15 different elements: H, He, C, N, O, Ne, Na, Mg, Al, Si, S, Ar, Ca, Fe, and Ni. Collisional ionization equilibrium (CIE) is assumed, such that the ionization state at a given temperature is given only by the balance of collisional ionization and recombination processes. 

The cooling rate is tabulated as a function of temperature such that $\Lambda_{\rm{net}} = \mathrm{d}(\rho u) / \mathrm{d}t = -n_e n_H \Lambda(T)$ is the internal energy per unit volume lost due to cooling. For practical purposes, the electron abundance $x_e = n_e/n_H$ is also tabulated, and used to express the cooling term in the Euler equations in terms of a normalized cooling rate \(\Lambda_\mathrm{Hydro} = x_e \Lambda\). Therefore, we take the tabulated values for \(\Lambda_{\mathrm{Hydro}}\) and add the term $\Lambda_{\rm{net}} = -n^2_H\Lambda_\mathrm{Hydro}$ to the energy conservation equation. We limit radiative cooling to gas with temperatures above \(10^4\,\mathrm{K}\), since the cooling function drops rapidly at temperatures below this threshold, and additional low-temperature physics becomes relevant.

\subsection{Turbulent Driving}

\begin{figure}
    \centering
    \includegraphics[trim={0 0 0 1cm},clip,width=0.49\textwidth]{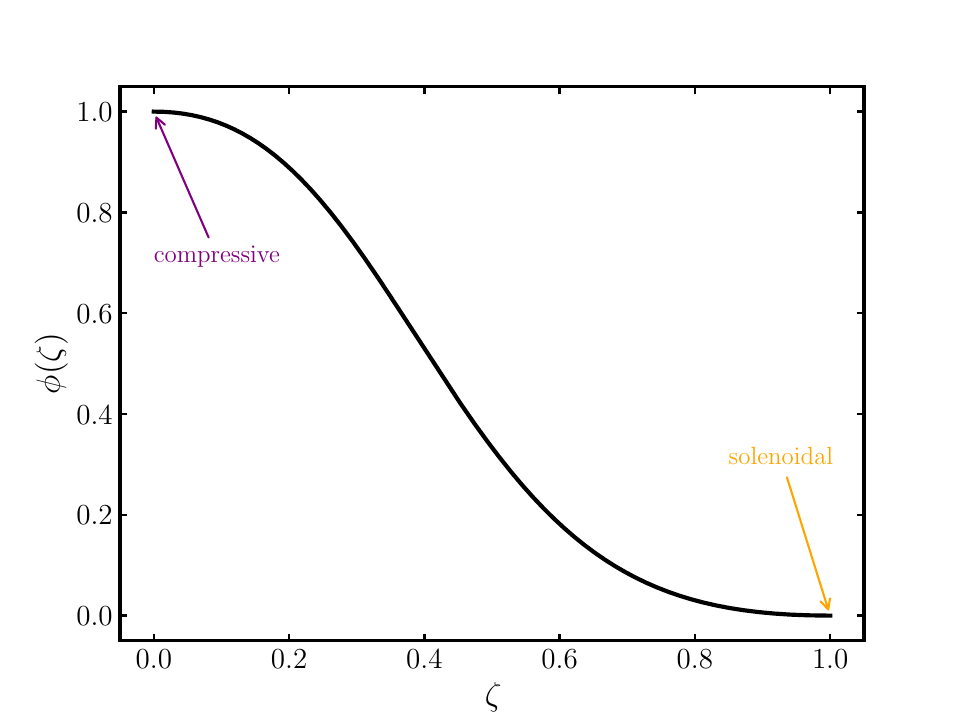}
    \caption{The ratio $\phi(\zeta)$ of compressive to total projection, as a function of the spectral weight $\zeta$. Intuitively, $\phi$ measures the ratio between the variance of the compressive component and the total after applying the spectral projection operator (see text). The ratio is zero for a solenoidal (divergence free) vector field and one for a vector field with only compressive component (rotation free).}
    \label{fig:forcingratio}
\end{figure}

To generate statistically stationary turbulent motions we constantly inject energy with an acceleration field only composed of low wave number (large-scale) components in Fourier space \citep{ESWARAN1988257}. A statistically stationary state of turbulence then develops as the energy cascades self consistently towards the smallest scales, where the rate of energy dissipation matches the rate of energy injection. 

To drive turbulent motions of the fluid, a stochastic force term $\bm{f}$ is introduced to the Euler equations, as noted above. We define the evolution of the driving force in Fourier space \citep{Bauer_2012}. In order to produce statistically homogeneous and isotropic turbulence, the driving force is evolved randomly for a discrete number of modes. In this way we can also specify the scale at which the driving force inputs energy into the system, which is important for quantifying the expected stationary turbulent flow \citep[following][]{2010A&A...512A..81F}. The implementation is heavily inspired by the \texttt{TurbGen} code \citep{2022ascl.soft04001F}. In particular, the random evolution is modeled by the Ornstein-Uhlenbeck (OU) process, which serves to control the amplitude and autocorrelation timescale of the forcing \citep{grete18}. Finally, using a Helmholtz decomposition operator in Fourier space, we can obtain purely divergence free (solenoidal) or compressive forcing fields, or a mixture of the two.

\subsubsection{Helmholtz Decomposition in Fourier space}

To decompose a vector field in its purely solenoidal and compressive components, we apply the projection operators $\mathcal{P}^{\parallel}_{ij}(\bm{k}) := k_i k_j/k^2$ and $\mathcal{P}^{\perp}_{ij}(\bm{k}) := \delta_{ij} - k_i k_j/k^2$ in Fourier space. Combining these two operators with the spectral weight $\zeta \in [0, 1]$:
\begin{align}
    \mathcal{P}^{\zeta}_{ij} := \zeta \mathcal{P}^{\perp}_{ij} + \left(1 - \zeta\right) \mathcal{P}^{\parallel}_{ij} = \zeta \delta_{ij} + \left(1 - 2\zeta\right) \frac{k_i k_j}{k^2}
\end{align}
forms the spectral decomposition operator. After applying the projection, a fully solenoidal force field ($\zeta = 1$), a fully compressive force field ($\zeta = 0$) or a mixture between the two are obtained. For example, applying the projection \(\mathcal{P}_{ij}^{\parallel}\) to the Fourier transform of the vector field \(\hat{\bm{f}}(\bm{k})\) has the following effect:
\begin{align}
    \nabla \times \int \mathcal{\bm{P}}(\bm{k})^{\parallel}\cdot\hat{\bm{f}}(\bm{k})e^{i\bm{k}\cdot\bm{x}}\mathrm{d}^3k = \int \bm{k}\times\frac{\hat{\bm{f}}(\bm{k})\cdot \bm{k}}{k^2}\bm{k}e^{i\bm{k}\cdot\bm{x}}\mathrm{d}^3k = 0,
\end{align}
and yields a vector field in real space with vanishing curl. To quantify the ratio between the compressive and total power, the (Frobenius) norms of the projection operators \citep{2010A&A...512A..81F} \(\Vert(1 - \zeta)P^{\parallel}\Vert^2 = (1 - \zeta)^2\) and \(\Vert P^{\zeta}\Vert^2 = 1 - 2\zeta + 3\zeta^2\) are combined into the forcing ratio:
\begin{align}
    \phi(\zeta) := \frac{(1 - \zeta)^2}{1 - 2\zeta + 3\zeta^2}
    \label{eq_forcing_ratio}
\end{align}
A more compressive driving force results in a more compressive velocity field. Figure \ref{fig:forcingratio} shows the forcing ratio $\phi(\zeta)$, illustrating how it gives a measure of compression. Note that we apply the spectral projection operator to a vector $\bm{f}$ of independent gaussian distributed components, such that its variance becomes $\langle (\mathcal{P}^{\zeta}(\bm{k})\cdot \bm{f})^2\rangle = \mathcal{P}^{\zeta}(\bm{k})_{ki} \mathcal{P}^{\zeta}(\bm{k})_{kj} \langle f_i f_j\rangle \sim \mathcal{P}^{\zeta}(\bm{k})_{ki} \mathcal{P}^{\zeta}(\bm{k})_{kj}\delta_{ij} = 1 - 2\zeta + 3\zeta^2$. The numerator of $\phi(\zeta)$ is found in a similar way. Therefore, it measures the ratio between the variance of the compressive component and the total after applying the projection. 

\subsubsection{Ornstein-Uhlenbeck process}

The OU process is a stochastic differential equation describing the evolution of the driving force in Fourier space for a given $\bm{k}$:
\begin{align}
    \mathrm{d}\hat{\bm{f}}(\bm{k}, t) = \sqrt{\frac{2}{T}}\sigma f_0(\bm{k})\mathcal{P}^\zeta(\bm{k})\cdot\mathrm{d}\bm{W}(t) - \hat{\bm{f}}(\bm{k}, t)\frac{\mathrm{d}t}{T}.
\end{align}
The right-hand side consists of a diffusion and a decay term, both involving the autocorrelation timescale $T$. The amplitude of the diffusion term is set by the overall scale $\bm{\sigma}$ and the dimensionless function $f_0(\bm{k})$, which specifies the relative magnitude between modes in Fourier space. Overall, the force depends on the choice of the parameters $\zeta$, $\sigma$ and $T$ as well as the shape of $f_0(\bm{k})$. The Wiener process (also called Brownian motion) is a random process defined by:
\begin{align}
    \mathrm{d}\bm{W}(t) = \bm{W}(t) - \bm{W}(t - dt) = \bm{N}(0, \mathrm{d}t),
\end{align}
where $\bm{N}(0, dt)$ is a vector of three independent Gaussian random variables with zero mean and variance $\mathrm{d}t$. The equation above is stochastic calculus notation for a random walk in three dimensions. The vector \(\bm{W}(t)\) changes independently of its past values. The separation of positions at times \(t\) and \(t^\prime\) is therefore random and follows a normal distribution \(\bm{W}(t^\prime) - \bm{W}(t) \sim \mathcal{N}(0, t^\prime - t)\).

Importantly, the OU process gives a random force field with decaying autocorrelation but statistically stationary amplitude. As a result, we use it to produce a stationary, statistically isotropic turbulent flow \citep{book}.

\subsubsection{Setting the force}

In practice, the driving force field must be discretized in space and time. The discretization in time is based on the solution of the stochastic differential equation of the OU process. Given a uniform time step $\Delta t$, the evolution of \(\hat{\bm{f}}^{(n)}(\bm{k}) = \hat{\bm{f}}(\bm{k}, t = n\Delta t)\) is determined by the sequence:
\begin{align}
    \hat{\bm{f}}(\bm{k}, t + \Delta t) =& e^{-\Delta t/T}\hat{\bm{f}}(\bm{k}, t) + \\& \sigma f_0(\bm{k})\sqrt{1 - e^{-2\Delta t/T}}\mathcal{P}^{\zeta}(\bm{k})\bm{z}_n, \nonumber
\end{align}
with $\bm{z}_n$ a vector of three independent gaussian normal distributed variables with zero mean and unit variance.

In this calculation we use the discrete Fourier transform to obtain the force in real space. This means that the driving force \(\hat{\bm{f}}(\bm{k}, t)\) is defined on a uniform grid in Fourier space (the spacing of the grid is given by the lowest wave number \(2\pi/L_\mathrm{box}\)). To not contaminate modes of higher wave numbers in the velocity field by the forcing, the amplitudes \(f_0(\bm{k})\) are nonzero only for a small range of low wave numbers. In practice, the amplitudes $f_0(\bm{k})$ are set up in a small interval of wave numbers $1 \cdot 2\pi/L_{\rm{box}} = k_{\mathrm{min}} < |\bm{k}| < k_{\mathrm{max}} = 3 \cdot 2\pi/L_{\rm{box}}$ and peaking at $k_{\mathrm{driv}}L_{\rm{box}}/2\pi = \frac{k_{\mathrm{min}} + k_{\mathrm{max}}}{2} = 2 \cdot 2\pi/L_{\rm{box}}$ \citep{2010A&A...512A..81F}. For this we define the amplitudes as:
\begin{align}
    f_0(\bm{k}) = \begin{cases} 
      \sqrt{1 -4\left(\frac{|\bm{k}| - k_{\mathrm{driv}}}{k_{\mathrm{max}} - k_{\mathrm{min}}}\right)^2}\frac{k_{driv}}{|\bm{k}|} & k_{\mathrm{min}} < |\bm{k}| < k_{\mathrm{max}} \\
      0 & \mathrm{else}.
   \end{cases}
\end{align}

\begin{figure}
    \centering
    \includegraphics[width=0.49\textwidth]{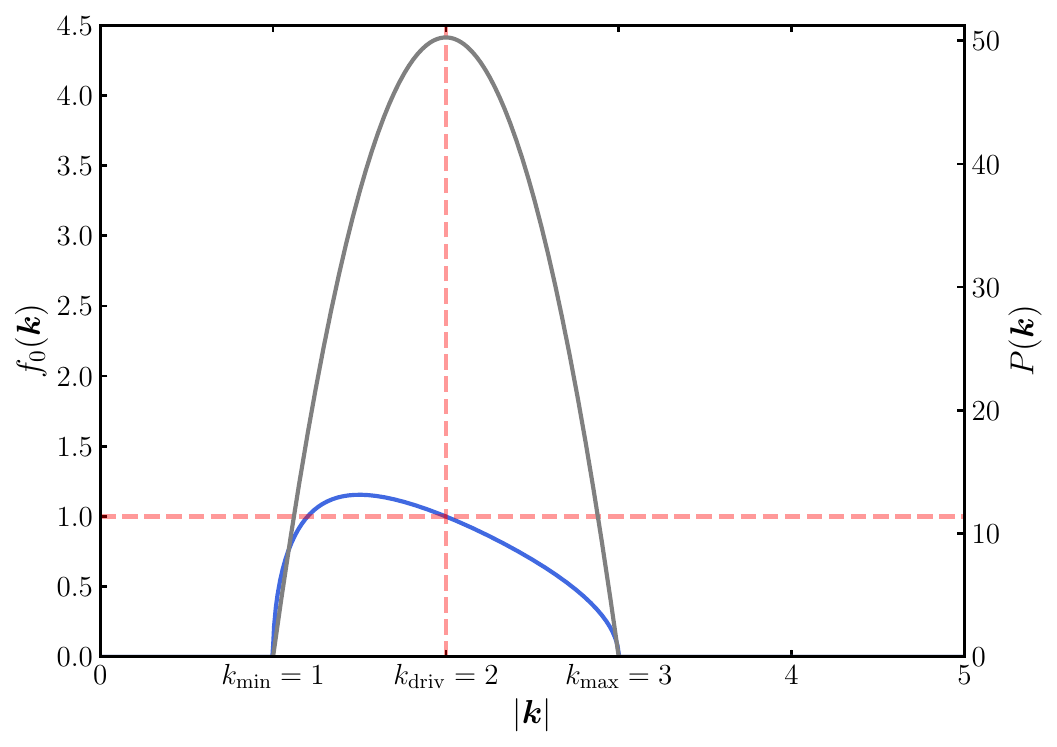}
    \caption{The driving amplitude $f_0(\bm{k})$ (blue) and the shape of the resulting power spectrum (gray), as a function of wave number $k$. The wavenumber at which the forcing power spectrum peaks effectively sets the scale of the inertial range of turbulent motions. Its value is chosen as a compromise between minimizing the influence of the periodic boundary conditions and retaining a sufficiently large inertial range for the turbulent cascade. Using a broader forcing profile centered around a characteristic wavenumber, rather than a delta function at a single mode, excites more independent Fourier modes and reduces artifacts caused by the discretization in Fourier space. Motions at higher wavenumbers are not contaminated by the forcing as a result of the amplitudes being identically zero at wavenumbers larger than $k_{\rm{max}}$. }
    \label{fig:driving_amplitude_f0}
\end{figure}

Figure \ref{fig:driving_amplitude_f0} shows the resulting 1D power spectrum, which is a parabola peaking at $k_{driv}$:
\begin{align}
    P(|\bm{k}|) = 4\pi|\bm{k}|^2\langle f_0(\bm{k})^2\rangle_{|\bm{k}|} \propto |\bm{k}|^2.
\end{align}
Since the physical driving field is not necessarily symmetric, its Fourier transform has real and imaginary parts in general. The total force is therefore a complex number, where both real and imaginary parts are given by independent Ornstein-Uhlenbeck processes:
\begin{align}
    \hat{\bm{f}}(\bm{k}) = \hat{\bm{a}}(\bm{k}) + i\hat{\bm{b}}(\bm{k})
\end{align} 
Finally, to obtain the force at a position $\bm{x}$ in real space, we take the real part of the discrete inverse Fourier transform:
\begin{align}
    \bm{f}(\bm{x}) &= \Re\left(\sum_{\bm{k}}\hat{\bm{f}}(\bm{k})e^{i\bm{k}\cdot\bm{x}}\right) \\
    &= \sum_{\bm{k}}\hat{\bm{a}}(\bm{k})\cos\left(\bm{k}\cdot\bm{x}\right) - \hat{\bm{b}}(\bm{k})\sin\left(\bm{k}\cdot\bm{x}\right) \nonumber
\end{align}
This calculation is free of resolution limits and is evaluated for every fluid cell in the simulation. 

\begin{figure}
    \centering
    \includegraphics[width=0.47\textwidth]{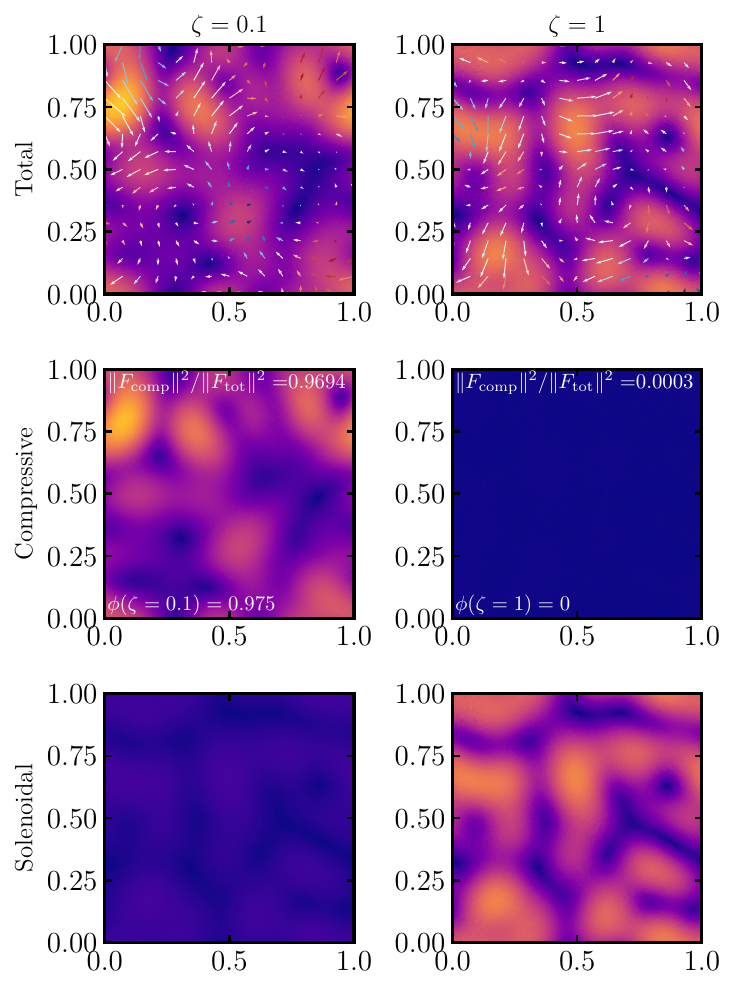}
    \caption{The driving force obtained in real space after an evolution of five autocorrelation times of the OU process in Fourier space. The columns show different realizations of the force for a spectral parameter \(\zeta = 1\) (solenoidal) and \(\zeta = 0.1\) (mostly compressive). The six panels show a 2d slice through the force generated in a 3d box with side length equal to one. The color represents the relative magnitude of the force field and is kept at a fixed range across the rows. The decomposition in compressive and solenoidal components is described in the Appendix.}
    \label{fig:driving-force}
\end{figure}

Figure \ref{fig:driving-force} shows two different example force fields obtained with this method. For the left column, the spectral weight in the decomposition is \(\zeta = 0.1\), meaning the force is a combination of mostly compressive and little solenoidal components. The right column shows a calculation of the force with \(\zeta = 1\), such that the force should be entirely solenoidal. The OU processes have a vanishing initial condition and are evolved for five autocorrelation times. The random number generator used for generating the normal distributed random variables starts with the same random seed. The decomposition of the force field in its compressive and solenoidal component is displayed in the panels below, showing only their magnitudes. This demonstrates that adjusting the parameter \(\zeta\) lets one obtain the wanted composition of the force (see Appendix for details). Comparing the ratio of the compressive to total power with the analytic forcing ratio \(\phi(\zeta)\) from (\ref{eq_forcing_ratio}) gives rough agreement, as shown in the middle row. Small differences arise because of the discretization errors in space. As can be seen in the third row, the solenoidal components have the same structure but appear with different magnitudes, as intended due to the two different spectral parameters.

\subsection{Multi-scale Turbulent Decomposition}

\begin{figure*}
    \centering
    \includegraphics[width=0.9\textwidth]{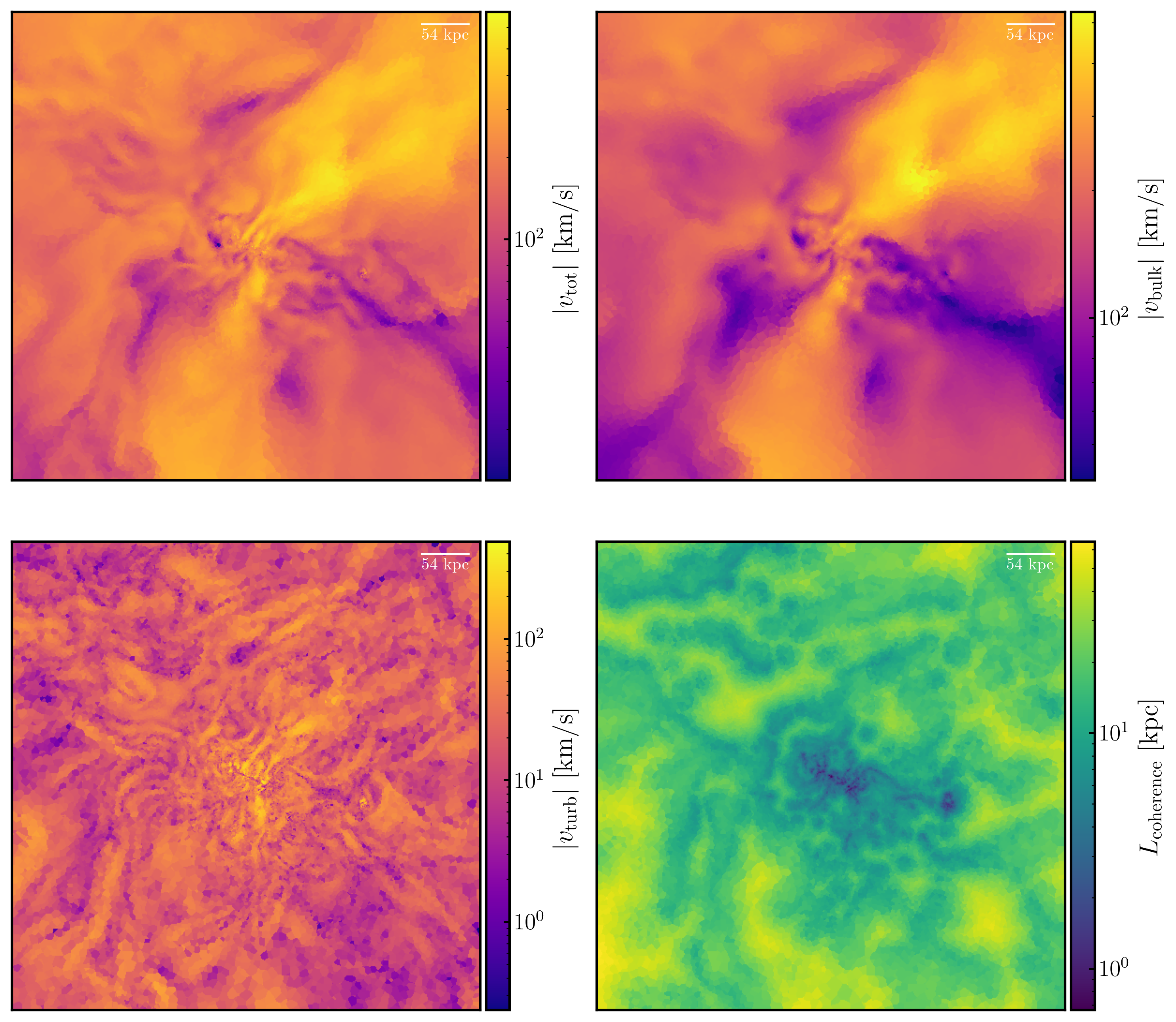}
    \caption{Visualization of the multi-scale velocity decomposition. We show a slice of total velocity magnitude (upper left), turbulent velocity (lower left), bulk velocity (upper right), and coherence length (lower right), i.e. the scale at which convergence is reached during the iterative method. The scale of the images contains the entire halo, out to the virial radius, for a randomly selected galaxy of the Milky Way/Andromeda-like galaxy catalog of TNG50 at $z=0$.}
    \label{fig:vel-turb-decomp}
\end{figure*}

We seek to distinguish large-scale gas characteristics from the fluctuations that are a result of turbulence. The decomposition in bulk (coherent) and turbulent (irregular or small scale) velocities and their scales is complicated by the fact that bulk motions occur across a variety of scales. Instead of defining a fixed spatial scale, we therefore apply a multi-scale filtering technique \citep{vazza2012,2021MNRAS.504..510V}.

In short, the flow structure above the scale of the largest turbulent motions (the integral scale) is uncorrelated, i.e. not governed by a turbulent cascade and the average velocity within this scale should tend to the average fluid velocity. By evaluating the average velocity of the flow in increasing spatial regions around a point, the integral scale can be identified as the scale at which the computed average velocity converges to some constant, the local velocity of the bulk flow. Therefore, our goal is to find the coherence scale \(L(\bm{x})\) for every position represented by a Voronoi cell center \(\bm{x}\). The final bulk velocity \(\left<\bm{v}\right>\) is then calculated as the mass weighted average of the velocities of the Voronoi cells contained in the sphere around \(\bm{x}\) with radius \(L(\bm{x})\). 

In particular, we first define the linear cell size \(\Delta x\) as the radius of a volume-equivalent sphere. We then initialize \(L_0 = 3\Delta x\), and calculate the mass weighted bulk velocity at the current coherence length \(L_n(\bm{x})\):
\begin{align}
    \left<\bm{v}\right>_n = \frac{\sum_{V(L_n)} m_i\bm{v}_i}{\sum_{V(L_n)} m_i},
\end{align}
where the sum is over all cells with cell center inside \(V(L_n)\), a sphere with radius \(L_n\) around \(\bm{x}\). The turbulent velocity is computed as \(\delta \bm{v}_n = \bm{v} - \left<\bm{v}\right>_n\). To assess convergence, we define a fractional change in \(\delta \bm{v}\) as:
\begin{align}
    \Delta = \max_{i = 1, 2, 3}\left|\frac{\delta v^i_n}{\delta v^i_{n - 1}} - 1\right|.
\end{align}
If this falls beyond a fixed tolerance parameter \(\Delta_{tol} = 0.05\), the iteration is stopped. That is, if the maximum of the relative change of every component of the turbulent velocity falls under some tolerance threshold. Alternatively, we also stop if a cell included in the computation of the bulk velocity contains a hydrodynamical shock, i.e. if the Mach number is above a threshold: \(\mathcal{M} > 1.3\). The shock Mach numbers are taken directly from the on-the-fly shock finder data of TNG50 \citep{Schaal_2014}. This is motivated by the fact that shocks introduce velocity jumps that highly bias the bulk velocity. If no stopping condition is triggered, we increase the outer scale by: $L_{n + 1} = \max\left(L_n + \Delta x, \left[1 + \chi\right]L_n\right)$ with \(\chi = 0.05\) fixed, and iterate. This is to prevent a slow convergence when a cell with small linear size has a large \(L(\bm{x}) \gg \Delta x\).

The method calculates a turbulent velocity vector and coherence length for each gas cell, from which the bulk velocity is derived as the difference between total and turbulent velocity. Since the iteration starts at a size comparable to the linear cell size, a lower limit for the coherence scale is set by the resolution.\footnote{We assess the dependence of the coherence scale on the spatial resolution (linear size) of our gas cells (not explicitly shown). We find a clear linear relation, where the calculated coherence scale grows with worsening resolution. In TNG50, gas cells are (de)refined such that the mass in each cell remains approximately constant. As a result, the resolution strongly decreases with density and, consequently, with distance from the galactic center. Nonetheless, it is possible that the integral scales of turbulent motions also decrease with distance, as bulk flows can become more complex inside or just beyond the disk. More work is needed to fully understand the limitations of the multi-scale filtering technique for simulations with variable spatial resolution.} Furthermore, we commonly terminate the method, in roughly 95\% of cases, because gas a shock Mach number larger than 1.3 enters the domain for bulk velocity calculation. In this case, the inferred coherence scale actually represents the distance to the nearest influential shock. \citet{vazza2012} argue that this condition is necessary since hydrodynamical shocks introduce a large skewness in the velocity distribution. This would slow down the convergence significantly and overestimate the scale of coherence.

\begin{figure*}
    \centering
    \includegraphics[width=0.99\textwidth]{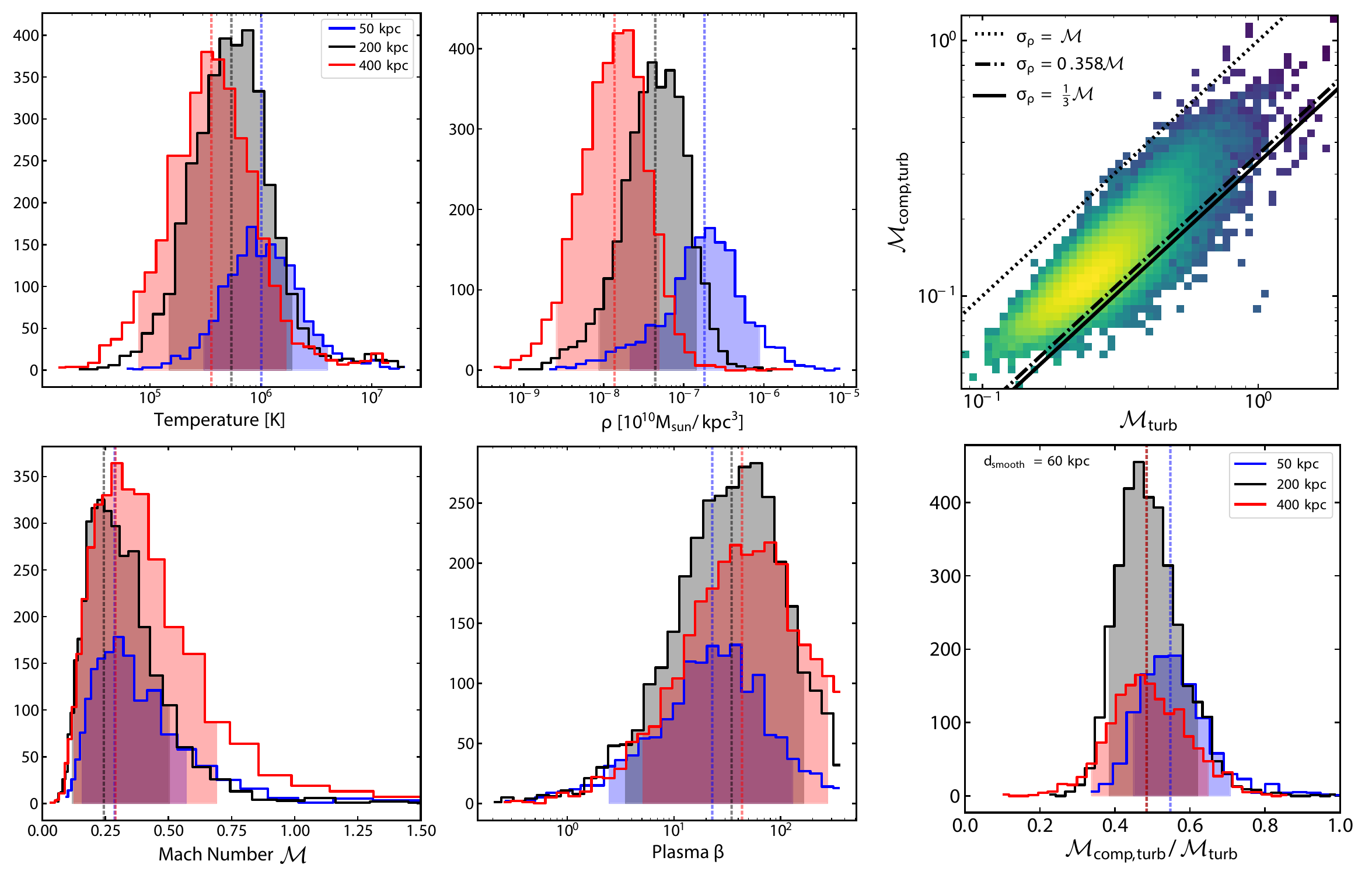}
    \caption{The distribution of CGM properties obtained using the smallest smoothing scale (15 kpc) is shown, grouped according to the different regimes. The ISM--CGM interface is indicated in blue, the CGM in black, and the CGM--IGM interface in red. The right two panels show the joint distribution of the Mach numbers based on the decomposed turbulent velocity field and its compressive component. Simulations of turbulence in idealized settings suggest that the ratio of compressive to total Mach number should lie between \(1/3\) for purely solenoidal and \(1\) for purely compressive driven turbulence. }
    \label{fig:cgm-props}
\end{figure*} 

We assess the impact of changing the Mach number threshold of 1.3, testing values of 1.1 and 1.5 (not explicitly shown). Across the majority of the CGM, this has little impact. However, in localized regions the coherence length is larger for a higher Mach number threshold and smaller for a lower threshold. The summed norm of the difference between the resulting turbulent velocity fields gives a difference of $\sim 9$\% for the higher ($\mathcal{M} = 1.5$) threshold and $\sim 4$\% for the lower threshold.


\section{Results: Milky Way-like Galaxies in TNG50} \label{sec_results1}

{\renewcommand{\arraystretch}{1.5}
\begin{table*}
    \centering
    \begin{tabular}{cccc}
    \textbf{Physical property} & \textbf{Disk-CGM Interface} & \textbf{CGM Region} & \textbf{CGM-IGM Interface} \\
    \hline\hline
    \(\rho~[M_\odot/\mathrm{kpc}^3]\) 
      & \(1.81 \cdot 10^{3}~[1.63 \cdot 10^{2} - 1.09 \cdot 10^{4}]\) 
      & \(4.40 \cdot 10^{2}~[7.51 \cdot 10^{1} - 1.70 \cdot 10^{3}]\) 
      & \(1.37 \cdot 10^{2}~[2.51 \cdot 10^{1} - 5.65 \cdot 10^{2}]\) \\
    \(T~[\mathrm{K}]\) 
      & \(1.01 \cdot 10^{6}~[2.80 \cdot 10^{5} - 4.22 \cdot 10^{6}]\) 
      & \(5.43 \cdot 10^{5}~[1.40 \cdot 10^{5} - 2.13 \cdot 10^{6}]\) 
      & \(3.59 \cdot 10^{5}~[7.44 \cdot 10^{4} - 1.73 \cdot 10^{6}]\) \\
    \(\mathcal{M}_{\mathrm{tot}}\) 
      & \(0.287~[0.143 - 0.644]\) 
      & \(0.244~[0.115 - 0.518]\) 
      & \(0.290~[0.123 - 0.708]\) \\
    \(\beta_{\mathrm{plasma}}\) 
      & \(22.8~[2.23 - 147]\) 
      & \(34.7~[3.44 - 204]\) 
      & \(43.4~[4.18 - 276]\) \\
    \(b = \mathcal{M}_{\mathrm{comp}}/\mathcal{M}_{\mathrm{tot}}\) 
      & \(0.548~[0.424 - 0.736]\) 
      & \(0.485~[0.373 - 0.645]\) 
      & \(0.484~[0.329 - 0.667]\) \\
    \hline
    \end{tabular}
    \caption{Physical quantities in three distinct regions that we adopt as the initial conditions of our idealized turbulent boxes. The three columns show the disk–CGM Interface (\(0 - 100\) kpc), the CGM (\(100 - 300\) kpc), and the CGM–IGM Interface (\(> 300\) kpc). The quantities are: gas density $\rho$, temperature $T$, turbulent Mach number $\mathcal{M}$, plasma $\beta$, and compressive ratio $b$.}
    \label{table:ics}
\end{table*}}

We analyze the selection of Milky Way (MW) and Andromeda (M31) like galaxies in TNG50 \citep{2024MNRAS.535.1721P} with the goal of extracting a reasonable parameter space for idealized simulations of turbulence. To begin, Figure \ref{fig:vel-turb-decomp} visualizes the results of the multi-scale turbulent decomposition method applied to a Milky-Way like galaxy at $z=0$ from TNG50. The four panels all show the same, halo-scale view of gas in the circumgalactic medium: the magnitude of the total velocity field (upper left), the decomposed bulk velocity field (upper right), turbulent velocity field (lower left), and inferred coherence scale (lower right).

The central SMBH of this galaxy is in the kinetic feedback mode, and is driving a large-scale galactic outflow perpendicular to the galactic plane (roughly parallel to the diagonal from upper left to lower right). This flow is prominently visible in the bulk velocity, whereas the turbulent velocity component more uniformly fills the circumgalactic medium. The latter has lower velocities, characteristic of subsonic motion, and coherence scales of $\sim \rm{1-10s}$ of kpc.

We characterize basic thermodynamical quantities including gas density, temperature, and metallicity, as well as the strength and character of turbulent motions. We measure the r.m.s. sonic Mach number (hereafter, just Mach number), a dimensionless number characterizing compressive turbulence, as $\mathcal{M} = u / c_s$, the fluid velocity magnitude normalized to local sound speed. We also decompose the velocity fields into turbulent and bulk components. If \(\bm{v}_\mathrm{turb}\) is the calculated turbulent velocity field, the turbulent Mach number is:
\begin{align}
    \mathcal{M}_\mathrm{turb} = \sqrt{\langle\bm{v}^2_\mathrm{turb} / c^2_s\rangle}.
    \label{eq:turbmach}
\end{align}
The turbulent velocity can also be split into compressive and solenoidal component via a Helmholtz decomposition: \(\bm{v}_\mathrm{turb} = \bm{v}_\mathrm{turb, comp} + \bm{v}_\mathrm{turb, sol}\) with \(\nabla \times \bm{v}_\mathrm{turb, comp} = 0\) and \(\nabla \cdot \bm{v}_\mathrm{turb, sol} = 0\) (see Appendix). The compressive parameter \(b\) quantifies the extent to which the driving injects compressive motions. Motivated by the relationship between density fluctuations and turbulent Mach number in Eqn. \ref{eq:bwithcompmach}, we estimate it as the ratio of compressive to total turbulent Mach number \(b = \mathcal{M}_\mathrm{turb, comp}/\mathcal{M}_\mathrm{turb}\). This is clearly a simplification, and assumes that energy injection occurs at a fixed scale and that the turbulent Mach number is computed over the corresponding bulk flow scale. In the CGM, turbulent motions may be driven by multiple processes operating at different scales, each with varying degrees of compressiveness.

To account for the global geometry of the CGM, these values are measured at different locations. We compute local averages of quantities like gas temperature by considering small cubic regions. Any region containing star-forming gas, rare in the CGM, is omitted. For each of 50 galaxies in the MW-M31 sample, we place $10^3$ such cubic sub-regions uniformly across the halo, adopting an averaging (i.e. cube) size of \(15 \,\mathrm{kpc}^3\) as our fiducial choice.\footnote{The Mach number shows a clear dependence on the smoothing length, while the compressive Mach number is nearly independent. This dependence could either reflect a true decline of turbulent Mach numbers on smaller scales, and/or numerical resolution effects. Note that the average number of Voronoi cells in a cutout of the MWM31 catalog is \(\sim 1.8\times10^6\). For a uniform distribution, a cube of volume \(60 \,\mathrm{kpc}^3\) contains about \(2500\) cells, while a \(15 \,\mathrm{kpc}^3\) cube contains only about \(40\) on average. A Helmholtz decomposition of the velocity field can still be performed in cubes with as few as \(40\) cells, since the Voronoi grid defines the velocity everywhere in space. However, in regions of low resolution, discontinuities at cell boundaries can introduce systematic errors in the decomposition. }

To understand the importance of distance from the central galaxy, we consider three radial regimes. For context, the galactic disks in our sample extend to $\sim 15$ kpc, while the average virial radius is $\sim 350$ kpc, roughly marking the outer edge of the CGM. We therefore define and label three distinct regions: (i) within 100 kpc of the galactic center as the ISM-CGM interface, (ii) from 100 to 300 kpc as the CGM, and (iii) beyond 300 kpc as the CGM-IGM interface.

Figure \ref{fig:cgm-props} examines the distributions of our quantities of interest in each of these three radial regimes (blue, gray, and red, respectively). Temperature (upper left) monotonically decreases with distance, as does density (upper center). Gas density distributions are broad with significant scatter even at fixed radii, hinting at the insufficiency of spherically averaged radial profiles \citep{dutta24}. Mach numbers are $<1$ at all distances, while plasma $\beta \sim 10-100$, indicative of thermal pressure dominance. Both have only weak trends with galactocentric distance. Note that the relative importance of turbulence in the CGM, as well as the existence of supersonic turbulence in the halo gas, may be significantly different at high redshift \citep{kakoly25}.

To quantify the nature of the turbulence, Figure \ref{fig:cgm-props} also shows the marginalized and joint distributions of the compressive parameter for the driving force as the ratio of the compressive to the total Mach number (right column). From idealized simulations of turbulence, we expect these values to lie within a band, with $b$ ranging from $1/3$ to $1$ \citep{2010A&A...512A..81F}. The upper right panel shows this joint distribution for a smoothing length of 60 kpc, where we find this constraint is largely satisfied. At smaller smoothing scales, we find that significant volumes of CGM gas have $b < 1/3$ (not shown), below the expected lower threshold for purely solenoidal turbulence. We suspect that resolution effects strongly influence the measurement of the compressive Mach number at these scales, and therefore adopt $b$ values as shown. Overall, $\mathcal{M}_{\rm comp,turb} / \mathcal{M}_{\rm turb} \sim 0.5$, indicating a mix of compressive and solenoidal turbulent motion.

Table \ref{table:ics} summarizes the results, providing the median and 5th to 95th percentiles for the values calculated with 15 kpc side length,\footnote{Ideally, we would calculate gas properties averaged on 1\,kpc scales, to match the intended domain of our idealized simulations, but the TNG50 resolution is insufficient to do so.} with an exception for the measurement of the fraction of compressive and total Mach number. Here, several caveats exist. First, it only makes sense to measure the compressive Mach number at the driving scale. In addition, density - Mach number relationships in the literature apply to isothermal conditions, and do not include density stratification or similar complications. It is also possible that the turbulence present in the CGM is not stationary, as it could be rising or decaying due to e.g. recent feedback or merger activity. Nonetheless, the instantaneous state of the $z=0$ circumgalactic medium from these TNG50 galaxies provides a physically motivated parameter space for our idealized simulations. 


\section{Results: Idealized Turbulence Simulations} \label{sec_results2}

{\renewcommand{\arraystretch}{1.5}
\begin{table*}
    \centering
    \begin{tabular}{lccccccccc}
    \textbf{Simulation} & \textbf{\(\zeta\)} & \textbf{\(V\)} & \textbf{\(T = L_\mathrm{driv}/V\)} & \textbf{\(t_\mathrm{fin} = 10T\)} & \textbf{\(\alpha_\mathrm{corr}\)} & \textbf{\(\mathcal{M}\)} & \textbf{\(\sigma_{\rho/ \langle\rho\rangle}\)} & \textbf{\(\sigma_s\)} & \textbf{\(\mathcal{M}_\mathrm{comp}\)}\\
    \hline\hline
    iso0.5-sol & 1 & 37.8 = \(0.5c_s\) & 0.0132 & 0.132 & 2.5 & \(0.483 \pm 0.004\) & \(0.0075 \pm 0.003\) & \(0.075 \pm 0.003\) & \(0.056 \pm 0.003\)\\
    iso0.5-mix & 0.5 & 37.8 = \(0.5c_s\)& 0.0132 & 0.132 & 2.5 & \(0.429 \pm 0.002\) & \(0.0077 \pm 0.004\) & \(0.077 \pm 0.004\) & \(0.064 \pm 0.003\)\\
    iso0.5-comp & 0.1 & 37.8 = \(0.5c_s\) & 0.0132 & 0.132 & 10 & \(0.580 \pm 0.010\) & \(0.390 \pm 0.025\) & \(0.405 \pm 0.028\) & \(0.320 \pm 0.017\)\\
    iso10-sol & 1 & 756 = \(10c_s\) & 0.000661 & 0.00661 & 2.5 & \(7.19 \pm 0.26\) & \(1.28 \pm 0.03\) & \(2.05 \pm 0.09\) & \(3.296 \pm 0.132\)\\
    iso10-mix & 0.5 & 756 = \(10c_s\) & 0.000661 & 0.00661 & 2.5 & \(6.53 \pm 0.22\) & \(1.46 \pm 0.07\) & \(2.73 \pm 0.30\) & \(3.518 \pm 0.224\)\\
    \hline
    \end{tabular}
    \caption{Parameters for the turbulent driving of the five isothermal simulations (first five columns). The remaining columns show measurements, time-averaged during the late stationary state: the Mach number, the standard deviation of density \(\rho\), of density contrast \(s\), and the compressive Mach number.}
    \label{table:iso-simulations}
\end{table*}}

To test the turbulent driving we first run five simulations of gas with an isothermal equation of state \(p = c_s^2\rho\), permitting a direct comparison to the results of \citet{2010A&A...512A..81F}. In this work, statistical properties and the density variance Mach number relationship were investigated for driven isothermal. To do so, we employ the specialized Riemann solver for isothermal hydrodynamics within \texttt{Arepo}. 

We explore different driving parameters. The general setup is a three dimensional cube with an equal side length of unity (in code units) and periodic boundary conditions. The simulation starts with uniform initial conditions across the cube and the mesh generating points are placed equally spaced in a grid consisting of \(128^3\) points. We adopt a unit system based on 1\,kpc, 1\,$\rm{M}_\odot$, and 1\,$\rm{km s^{-1}}$, and in these units use test simulations with an initial density of 1000, zero velocity, and a sound speed of 75.6. The initial conditions and important parameters for the simulations are summarized in Table \ref{table:iso-simulations}. For isothermal simulations, the initial state is specified by the density, velocity and sound speed. Five different simulations are run to compare the behavior of the driving, subsonic and supersonic turbulence, and different ratios of compressive to total forcing. 

We set the autocorrelation timescale \(T\) equal to the large-eddy turn-over timescale $L/V$ with the desired value of the stationary velocity \(V\) and the driving scale \(L_\mathrm{driv} = L_\mathrm{box}/2\), such that \(T = L_\mathrm{driv}/V = 1/(2V)\). This gives a rate of kinetic energy injection $\dot{E} = V^2/T = V^3/L$. By dimensional analysis, a scale of the needed force $F$ per unit mass based on the rate of injected kinetic energy can be constructed as $F^2 = \dot{E}/T$. Therefore, we set the variance $\sigma^2$ of the OU process to be:
\begin{align}
    \sigma^2 = \frac{E_\mathrm{corr}}{T} = \frac{(\alpha_\mathrm{corr}V)^3}{L_\mathrm{driv}T},
\end{align}
as in the description of the method above, but with a factor \(\alpha_\mathrm{corr}\) to adjust the energy injection quantitatively \citep{2022ascl.soft04001F}. This is important to give exact control over the actual stationary velocity. The driving has variable effectiveness for different Mach numbers and spectral parameters. The factor \(\alpha\) is also included in the same way in the \texttt{TurbGen} code. It is not clear a priori in what way this correction factor depends on \(\zeta\) or \(V\) and for the purposes of the simulations, \(\alpha_\mathrm{corr}\) was determined in order to obtain the desired velocity. In reality, the actual injection of energy likely depends on the nature of the forcing. The total energy injection rate by the turbulent driving force \(f\) is \citep{Mohapatra_2021}:
\begin{align}
    \int_{V_\mathrm{box}} \rho \bm{f}(\bm{x}) \cdot \bm{v}(\bm{x})\mathrm{d}V,
\end{align}
where \(\bm{v}\) is the fluid velocity. Note that the dot product in the integral can be negative, reducing the overall expected efficiency of the driving.

The simulations are run for ten autocorrelation times to ensure the formation of a statistically stationary state. The Mach number (in the isothermal case, this is the velocity normalized by the sound speed) stabilizes after roughly two autocorrelation times, indicating that the flow has reached a stationary state. At this point, the rate of numerical dissipation matches the rate of energy injection by the turbulent driving, such that the average kinetic energy per unit mass stays constant.

\begin{figure*}
    \centering
    \includegraphics[width=0.75\textwidth]{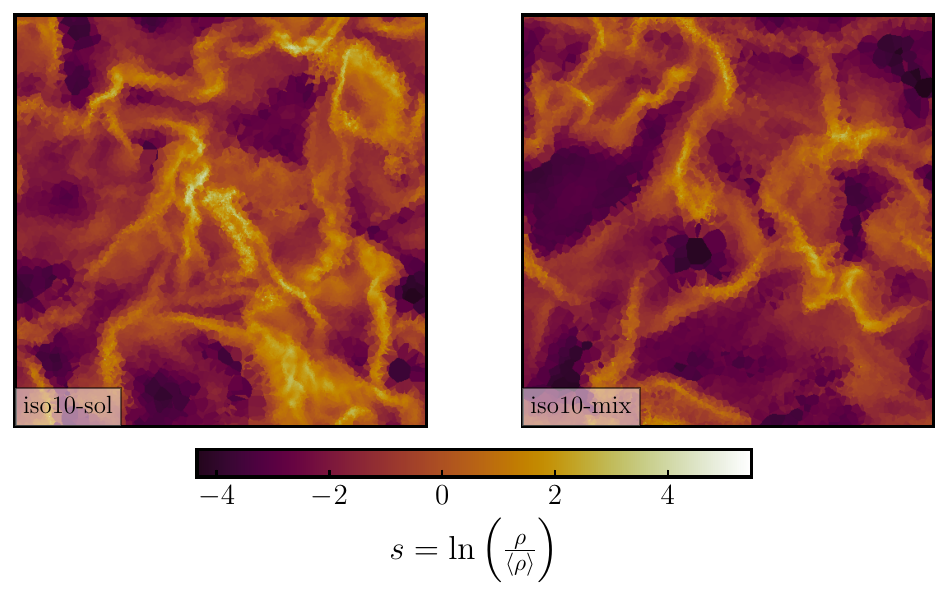}
    \includegraphics[width=0.9\textwidth]{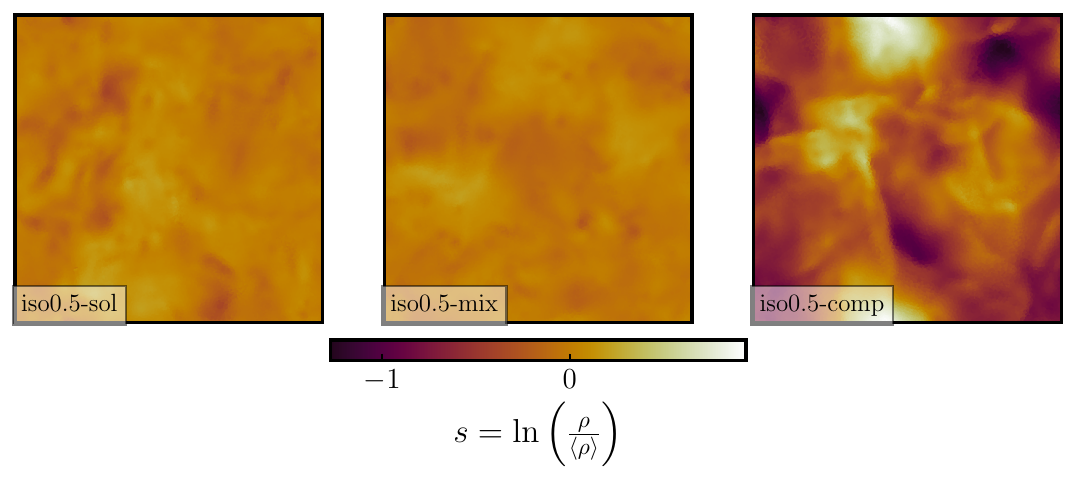}
    \caption{Visual comparison of the density contrast for supersonic (top row) and subsonic (bottom row) simulations. Shown are slices through the middle of the 3D simulation volume, at the end of each simulation. Compressive driving results in a higher absolute value of the density contrast, as expected. The supersonic simulations exhibit density contrasts several orders of magnitude higher than in the subsonic simulations (note the different colorbar ranges).}
    \label{fig:dens-image}
\end{figure*}

Figure \ref{fig:dens-image} shows slices of density at the end of each simulation. The local density contrast \(s = \ln\left(\rho / \langle\rho\rangle\right)\) measures the logarithmic deviation from the initial mean density \(\langle\rho\rangle\). The influence of the driving becomes clear, and we see that increasing either the Mach number or the compressive component creates density contrast values of higher absolute magnitude. 

\subsubsection{Density PDF-Mach number Relation}

The probability distribution function (PDF) of gas density in a turbulent medium is characterized by some level of universality, depending on the physics of turbulence \citep{Federrath_2012}. In the case of an isothermal equation of state, and assuming that the velocity divergence fluctuates randomly, \citet{PhysRevE.58.4501} argue that the density contrast $s$ follows a gaussian distribution:
\begin{align}
    p_s \mathrm{d}s = \frac{1}{\sqrt{2\pi\sigma_s^2}}\exp\left[-\frac{(s - \langle s\rangle)^2}{2\sigma_s^2}\right]\mathrm{d}s
\end{align}
The mean and variance of this distribution are related due to mass conservation: $\langle s\rangle = - \frac{1}{2}\sigma_s$, and the variance of the density PDF is related to \(\sigma_s\) with: $\sigma^2_s = \ln (1 + \sigma^2_{\rho/\rho_0} )$. \citet{2008ApJ...688L..79F} investigate a relationship between density dispersion and the root-mean-square Mach number \(\mathcal{M} = \sqrt{\langle \bm{u}^2/c^2_s \rangle_{\bm{x}}}\) in isothermal simulations of forced turbulence:
\begin{align}
    \frac{\sigma_\rho}{\rho_0} = b \mathcal{M},
    \label{eq:basic_densmach}
\end{align}
or equivalently \(\sigma^2_s = \ln\left(1 + b^2\mathcal{M}^2\right)\). In this case, density fluctuations scale with the Mach number of the turbulent flow. The parameter \(b\) is the compressive parameter and varies smoothly from \(b = 1/3\) for purely solenoidal (only rotational) to \(b = 1\) for purely compressive turbulent forcing. More precisely, \citet{2010A&A...512A..81F} find that for stochastically driven turbulence with accelerations \(F_\mathrm{tot}\),
\begin{align}
    b = \frac{1}{3} + \frac{2}{3}\left(\frac{F_\mathrm{comp}}{F_\mathrm{tot}}\right)^3,
    \label{eq:comb_forces}
\end{align}
where \(F_\mathrm{comp}\) is the total power of the compressive component of the acceleration field. As a result, compressive turbulent flows depend on the energy injection mechanism, especially on the whether compressive or solenoidal modes of the fluid motion are excited. 

Similarly, \citet{Konstandin_2012} show that replacing the Mach number in Eqn. \ref{eq:basic_densmach} by the Mach number formed with only the compressive component of the velocity \(\mathcal{M}_\mathrm{comp}\) provides a reasonable fit for the density dispersion Mach number relation. The compressive component of the velocity plays a role, since density variations occur due to compressive modes of the velocity field. They measure the relationship \(\sigma_s^2 = \ln\left(1 + \mathcal{M}^2_\mathrm{comp}\right)\) and therefore give implicitly another possible way to measure the compressive parameter as the fraction of compressive to total Mach number:
\begin{align}
    b = \frac{\mathcal{M}_\mathrm{comp}}{\mathcal{M}}
    \label{eq:bwithcompmach}
\end{align}
where \(\mathcal{M}_\mathrm{comp} = \sqrt{\langle \bm{u}_\mathrm{comp}^2/c^2_s \rangle_{\bm{x}}} = |\bm{\mathcal{M}}^2_\mathrm{comp}| = \sqrt{3}\mathcal{M}_\mathrm{comp, x/y/z}\) for isotropic compressive velocities. Importantly, strong deviations from this relation exist at very low and high Mach numbers. As with most results here, this finding is also based on driven isothermal turbulence.

\citet{2012MNRAS.423.2680M} provide a theoretical derivation for the density contrast variance Mach number relationship,

\begin{align}
    \sigma^2_s = \ln\left(1 + \frac{(\gamma + 1)b^2\mathcal{M}^2}{(\gamma - 1)b^2\mathcal{M}^2 + 2}\right).
\end{align}

Setting \(\gamma = 1\) results in the original relation for isothermal gas. 

Overall, the relationship between density fluctuations and the Mach number of the turbulent flow as well as the general PDF of the density is an ongoing topic of research. In particular, the combination of turbulence with magnetic fields, self-gravity, non-isothermal equations of state, or density stratification may all lead to non-trivial modifications of these results.

In order to measure the variance of density contrast, we must choose appropriate weights \(w_i\), such that the mean and variance of a sample of \(N\) data points \(Q_i\) is given by:
\begin{align}
    \langle Q \rangle = \frac{\sum_{i = 0}^{N}Q_i w_i}{\sum_{i = 0}^{N}w_i}\qquad\mathrm{and}\qquad \sigma^2_Q = \langle (Q - \langle Q\rangle)^2\rangle.
\end{align}
\citet{Konstandin_2012} argue that for the measurement of the density contrast variance \(\sigma^2_s\), the cell masses should be used as weights instead of the the cells volumes. If the PDF of \(s\) follows a log-normal distribution, the variance \(\sigma_s\) and the variance of the density PDF are related by: 
\begin{align}
    \sigma^2_s = \ln (1 + \sigma^2_{\rho/ \langle\rho\rangle}).
\end{align}
and that work showed that the mass weighted variance is more compatible with this relationship. In the results below, we therefore estimate the density contrast variance using mass weighting.

We also measure the compressive Mach number \(\mathcal{M}_\mathrm{comp}\) as the compressive velocity component normalized by the sound speed, using a Helmholtz decomposition. Finally, we calculate the driving parameter \(b\) in two ways, following the discussion above:
\begin{align}
    b_1 &= \frac{1}{\mathcal{M}}\sqrt{e^{\sigma^2_s} - 1} \\
    b_2 &= \frac{\mathcal{M}_\mathrm{comp}}{\mathcal{M}}.
\end{align}
\begin{figure}
    \centering
    \includegraphics[trim={0.5cm 0 0 1.2cm},clip,width=0.52\textwidth]{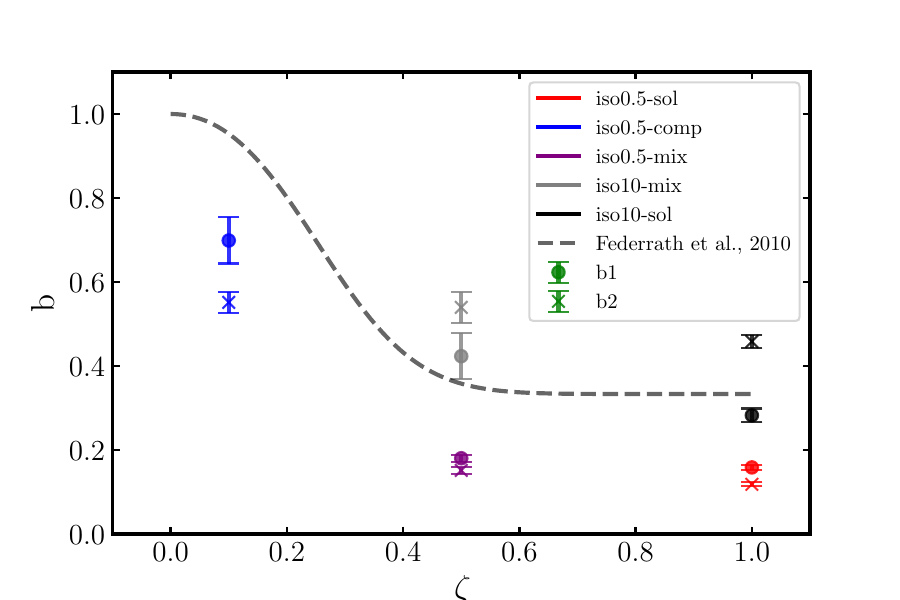}
    \caption{The compressive parameter \(b\) as a function of the spectral parameter \(\zeta\) that sets the compressiveness of the force. We include two distinct measurements $b_1$ (circles) and $b_2$ (crosses), as discussed in the text. The dashed line is an empirical fit obtained from simulations of supersonic turbulence with an isothermal equation of state \protect\citep{2010A&A...512A..81F} that is in broad agreement with our supersonic results.}
    \label{fig:zetavsb}
\end{figure}

Figure \ref{fig:zetavsb} shows our measured compressive parameters $b$ as a function of the spectral parameter $\zeta$. We show estimates ($b_1$ and $b_2$) as circles and crosses, respectively, with errorbars representing the variation in time.

\begin{figure*}
    \centering
    \includegraphics[width=0.86\textwidth]{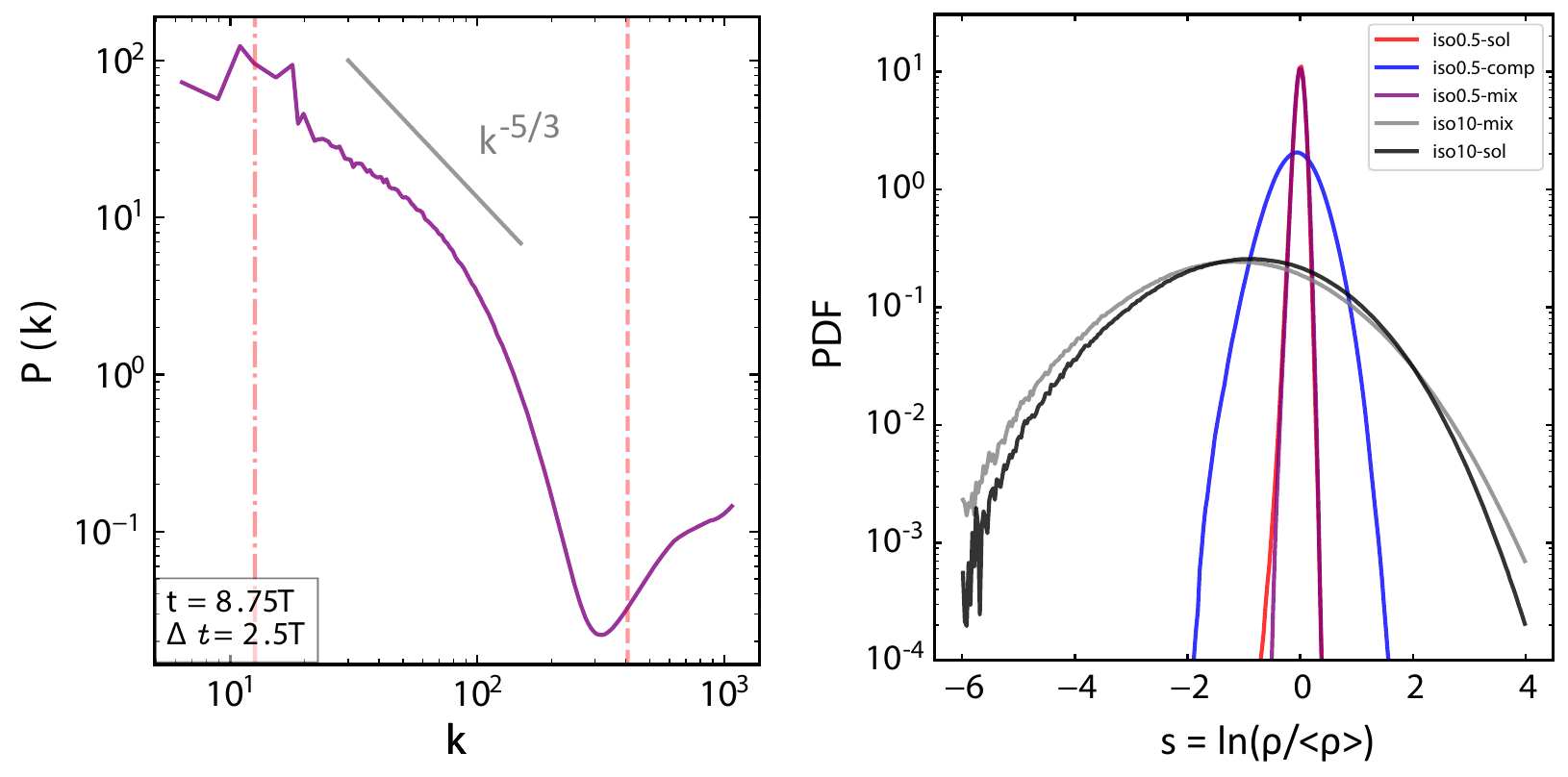}
    \caption{Left: The time-averaged one-dimensional velocity power spectrum at the end of the \texttt{iso0.5-sol} simulation. To show the steady state, we average over the past quarter of the entire simulation time. The grey line indicates a $-5/3$ powerlaw slope, while the dot-dashed and dashed red vertical lines indicate the smallest and largest gas cell sizes. Right: The volume weighted PDF of the density contrast \(s\) in all five simulations, averaged in the stationary regime of the flow over the last quarter of each simulation. The PDF is much broader for higher Mach numbers or more compressive driving.}
    \label{fig:powerspec-denspdf}
\end{figure*}

We compare our results to a fit from previous simulation results based on isothermal supersonic turbulence \citep[dashed line;][]{2010A&A...512A..81F},
\begin{align}
    \hat{b} = \frac{1}{3} + \frac{2}{3}\left(\frac{(1 - \zeta)^2}{1 - 2\zeta + 3\zeta^2}\right)^3.
    \label{eq:federrath-empirical-bzeta}
\end{align}
that agrees well for our supersonic simulations, as expected. However, in the subsonic range, the forcing is less effective at creating compression with the same spectral composition. One difference is that in the subsonic case, hydrodynamical shocks that drive sharp density contrasts are less frequent and the compressive component mostly drives sound waves. Our moderate numerical resolution may also somewhat hinder the formation of compressive motions on small scales. Finally, we note that the two methods of calculating \(b\) are only in rough agreement with each other, suggesting that the determination of the compressiveness requires caution.

\subsection{Velocity Power Spectrum}

Figure \ref{fig:powerspec-denspdf} shows the corresponding one-dimensional velocity power spectrum of the \texttt{iso0.5-sol} simulation at late time (left panel), averaged over the last quarter of the simulation. This is motivated by the fact that the statistically stationary state is reached, so noise in the power spectrum is reduced by averaging out local fluctuations in time. 

We measure the velocity power spectrum as follows (see Appendix for details): the three-dimensional power spectrum \(P(\bm{k})\) of a quantity \(\bm{v}(\bm{x})\) is proportional to the absolute value squared of its Fourier transform \(P(\bm{k}) \sim \vert \hat{\bm{v}}(\bm{k})\vert^2\). To calculate the Fourier transform of the velocity field, it first needs to be mapped to a cartesian grid on which we can calculate the discrete Fourier transform. A nearest neighbor sampling with the Voronoi cell generating points is then used to obtain the corresponding velocity values on this grid. Assuming the velocities are isotropic, the 1D power spectrum is then calculated by defining logarithmically spaced radial bins for \(k = \vert\bm{k}\vert\), to determine the angular averaged power spectrum: \(P(k) = 4\pi k^2\langle P(\bm{k})\rangle\).

This method can create noise for small scales because of sudden jumps in velocity at Voronoi cell boundaries. The red dashed line in Figure \ref{fig:powerspec-denspdf} marks the wave number corresponding to the linear cell size (\(r = V_\mathrm{cell}^{1/3}\)) of the largest cell in the simulation at that time (more precisely the maximum linear cell size during the averaging interval). Around this scale, the velocity statistics will be strongly influenced by sampling effects.

The time evolution shows the build-up of a stationary power spectrum with a few prominent features. The modes in which the driving interacts with the velocities can be seen as the large peak around the dash-dotted line, which marks the driving scale \(k_\mathrm{driv} = 2\pi/L_\mathrm{driv} = 4\pi/L_\mathrm{box}\). This peak is spread out since the driving also includes modes in a window around the driving scale. There is a small range from \(k = 20\) to \(k = 40\) that we identify as the inertial range, following a roughly constant slope. For modes with \(k > 100\) the power spectrum drops rapidly, indicating that the numerical dissipation dominates on these scales. In the identified inertial range, the power spectrum exhibits a power-law with a slope slightly shallower than expected by the Kolmogorov theory for incompressible turbulence.


\section{Results: Radiative Cooling} \label{sec_results3}

We now turn to idealized simulations of turbulence with the addition of energy loss via radiative cooling. This enables us to study the formation and development of a turbulent multi-phase gas. We run a suite of simulations that vary the driving strength, cooling function, computational mesh, and resolution. We set the spectral parameter \(\zeta\) to match the compressive parameter \(b = 0.548\), which is the median value of the disk-CGM interface. The value of \(\zeta\) is obtained with the fit function of \citet{2010A&A...512A..81F} as given by Eqn. (\ref{eq:federrath-empirical-bzeta}). As a result, the driving is a mixture of compressive and solenoidal modes with \(\zeta = 0.324\).

The initial density was picked as ten times the median density in the disk-CGM interface: \(1.8\times10^4\,\mathrm{M}_\odot/\mathrm{kpc}^3\), since for higher densities the cooling timescale is shorter, therefore facilitating the formation of cool and hot phases through radiative cooling. For reference, one simulation starts at ten times lower density, which also lets us assess the impact of this choice. The initial temperature is set to \(10^{5.5}\,\mathrm{K}\), since the cooling function increases sharply with decreasing temperature, allowing for a accelerated development of thermally unstable behavior. The amplitude of the driving is set such that the autocorrelation timescale is \(T = 11.17\,\mathrm{Myr}\) and the Mach number of gas with temperature above \(10^5\,\mathrm{K}\) approaches a value of roughly \(0.5\), which is in the range of the CGM Mach numbers. For one simulation, the driving amplitude was doubled to achieve a higher velocity and Mach number, to investigate the influence of stronger turbulence. Since the speed of sound of the gas depends on the temperature, we limit its calculation to the hotter phase (\(T > 10^5\,\mathrm{K}\)).

In total, we present six simulations, a fiducial setup as a baseline referred to as \texttt{gthmM0.5} and five additional runs for comparison. All simulations start with uniform initial temperature and density. They contain \(128^3\) resolution elements except for one higher resolution simulation with \(256^3\) resolution elements. The computation domain is a cube with side length of \(1\,\mathrm{kpc}\) with periodic boundary conditions, representing a small patch of gas in the CGM. One simulation is run with mesh movement disabled. Finally, we also include only primordial cooling in one simulation, instead of our fiducial metal-line cooling implementation. All simulations are run until \(t_\mathrm{end} = 10T = 111.7\,\mathrm{Myr}\).

Table \ref{table:cooling-sims} summarizes our runs, where the last column gives the average volume weighted Mach numbers of the hot phase (\(T > 10^5\,\mathrm{K}\)), averaged across the last 5 autocorrelation times, with errors showing the $1\sigma$ variation over this time. These are close (but not exactly equal) to the expected values, of either $\mathcal{M} = 0.5$ or $\mathcal{M} = 1$, as discussed above.

{\renewcommand{\arraystretch}{1.5}
\begin{table}
    \centering
    \begin{tabular}{lcccc}
    \hline
    \textbf{Name} & \textbf{$\rm{N_{\rm cells}}$} & \textbf{Mesh} & \textbf{Cooling} & \textbf{\(\mathcal{M}_\mathrm{hot}\)}\\
    \hline
    gthmM0.5 & \(128^3\) & moving & +metals & \(0.41 \pm 0.02\)\\
    gthmM0.5-static & \(128^3\) & static & +metals & \(0.49 \pm 0.03\)\\
    gthmM0.5-256 & \(256^3\) & moving & +metals & \(0.42 \pm 0.01\)\\
    gthmM0.5-primo & \(128^3\) & moving & primordial & \(0.56 \pm 0.01\)\\
    gthmM1 & \(128^3\) & moving & +metals & \(0.83 \pm 0.03\)\\
    gthmM0.5-lowdens & \(128^3\) & moving & +metals & \(0.57 \pm 0.02\)\\
    \hline
    \end{tabular}
    \caption{A summary of the setup and initial conditions of our idealized turbulence simulations that include cooling. All adopt the same $T_{\rm 0} = 10^{5.5}$\,K and $\rho_0 = 1.8 \times 10^4 \,\rm{M}_\odot/\rm{kpc}^3$.}
    \label{table:cooling-sims}
\end{table}}

We design the simulations such that they maintain global thermal energy balance, preventing a global, single phase cooling flow. Instead, only local thermal instabilities can develop. If the energy injected by the turbulent driving is not enough to counteract the total loss of energy due to cooling, the temperature of the gas would continue to drop without reaching any form of steady-state. Therefore, at every time step in the simulation we calculate the net energy added or lost as a combination of the turbulent driving and cooling. We add the calculated net energy uniformly back into the simulation. This leads to an additional heating term \(Q\) in the energy conservation equation with \(Q \propto \rho\).\footnote{This term is not guaranteed to be positive (heating). For example, if the energy injected by the turbulent dissipation is higher than the energy lost by radiative cooling at a given time step. Then, internal energy is actually subtracted uniformly across the simulation.} Physically, such a density dependent heating function could represent a background radiation field heating the gas. Our implementation is one choice, while other methods exist to enforce a global equilibrium. For example, \citet{Mohapatra_2021} adjust the driving amplitude based on the amount of energy lost by the cooling, achieving total energy balance.

\begin{figure*}
    \centering
    \includegraphics[width=0.9\textwidth]{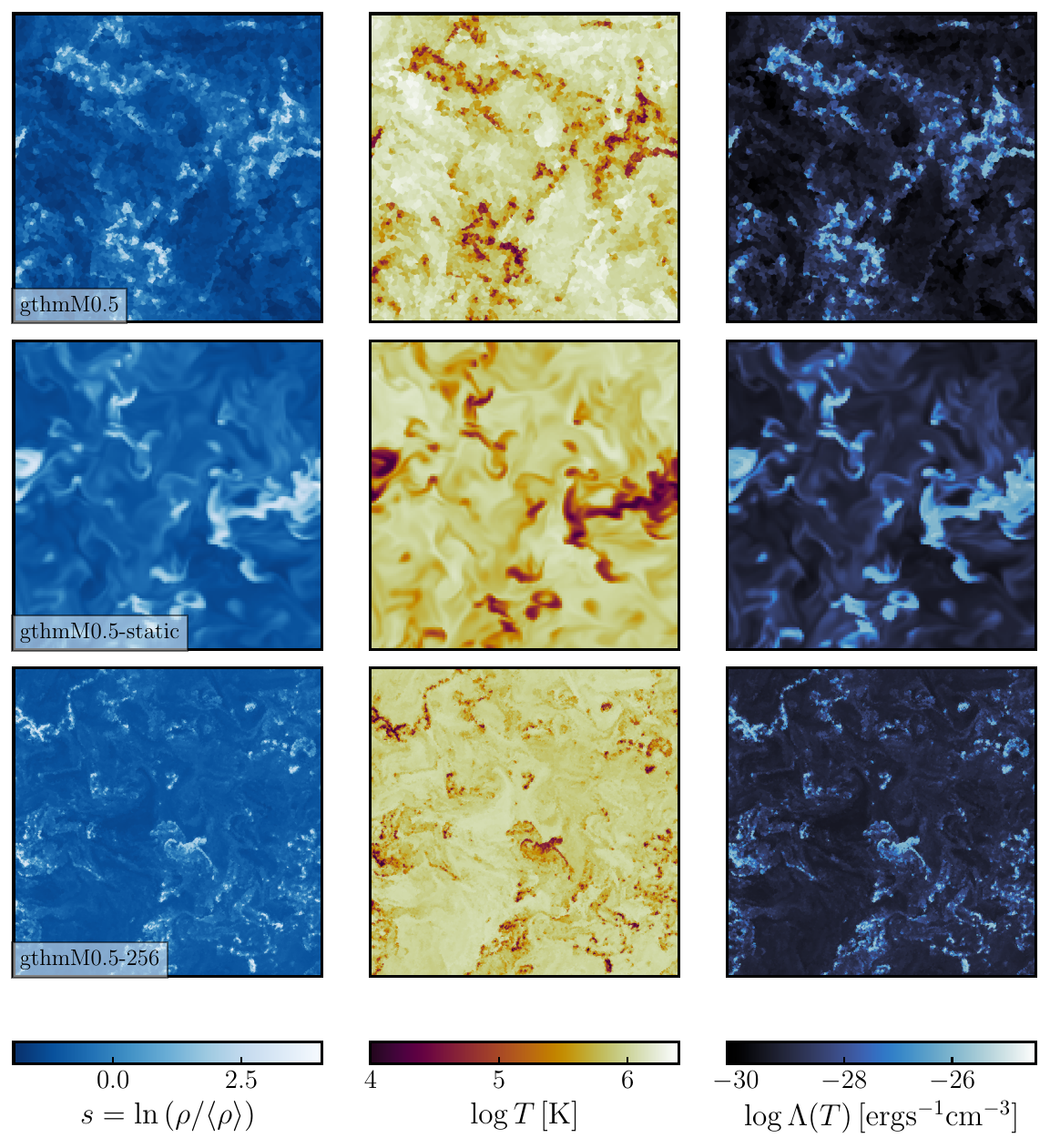}
    \caption{Slices through the simulation domain showing three different properties of the gas at the final snapshot: overdensity (left), temperature (center), and cooling rate (right). In comparison to a fixed mesh run with our fiducial parameters (center row), the moving-mesh (top) and higher numerical resolution (bottom) are both able to resolve significantly more small-scale, and cool, gas structure.}
    \label{fig:gthm-sims1}
\end{figure*}

Figure \ref{fig:gthm-sims1} visualizes slices of density contrast, temperature and volumetric cooling rate \(\Lambda(T) = n_en_H\Lambda_N(T)\) at the final snapshot of three of the simulations. We show our fiducial setup (top row), contrasted to the static mesh case (middle), and the higher resolution convergence test (bottom). These three simulations all show significant development of multiphase gas, with temperatures ranging from \(10^4\,\mathrm{K}\) to \(10^{6.5}\,\mathrm{K}\). The cool phase (\(T \sim 10^4\,\mathrm{K}\)) corresponds to regions of higher density and therefore higher cooling rates, and is distributed in small clumps and filaments, where as most of the volume is filled with hotter gas (\(T \sim 10^6\)).

In contrast, the two simulations with either primordial cooling only or lower initial density do not develop significant cool gas by $10T$, and exhibit much smaller density and temperature fluctuations (not shown).

\begin{figure*}
    \centering
    \includegraphics[width=0.98\textwidth]{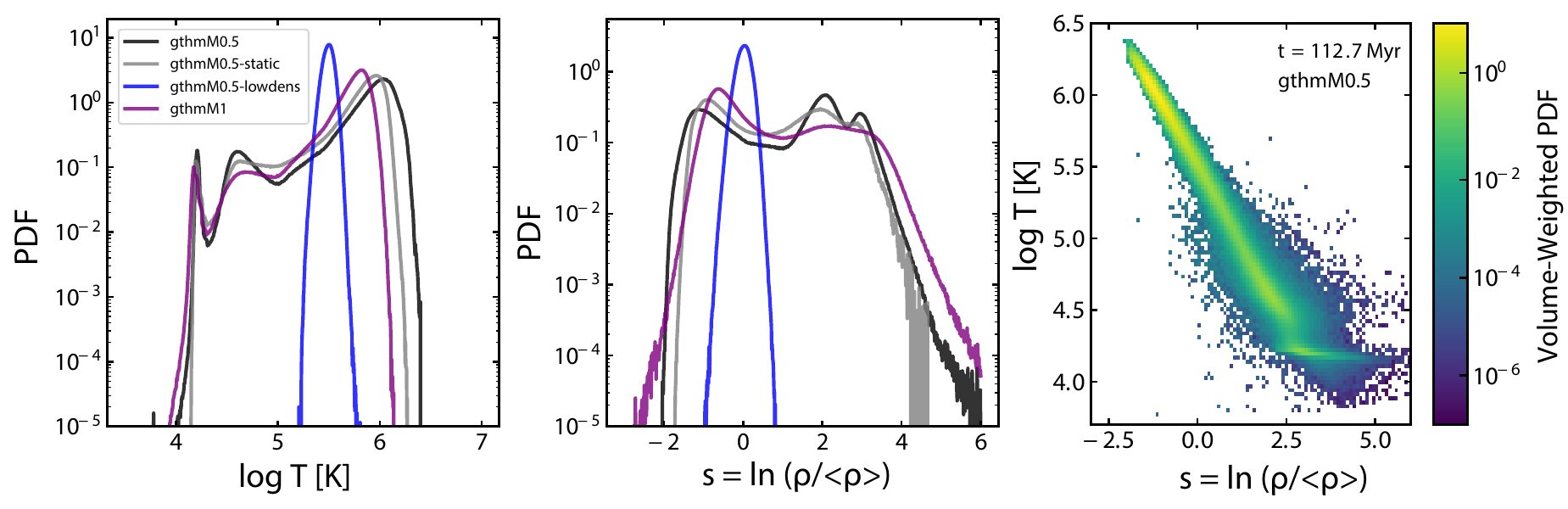}
    \caption{Volume weighted density contrast PDF (left), temperature PDF (center), and density-temperature phase space (right). For the first two we show all radiative cooling simulations, whereas for the last we focus on the final state of the \texttt{gthmM0.5} simulation as representative.}
    \label{fig:phase}
\end{figure*}

Figure \ref{fig:phase} shows the PDF of temperature and density contrast averaged over the last 10 snapshots of the simulation (left and center panels). We see that for the simulation starting with lower density, no multiphase medium develops, as the distribution of temperature and density is narrow around the initial value.\footnote{Figure \ref{fig:resolution-denstemppdf} of the Appendix demonstrates that these simulations, at least for the statistics shown, are well converged at our current numerical resolution.}

The distributions of the other simulations evolve markedly away from the initial conditions. They all follow roughly a similar qualitative shape, with peaks at high and low temperatures as well as densities. In addition, the shapes of the distributions in the case of moving and static mesh simulations show differences in the high and low density tails of the density PDF. This demonstrates the advantages of the spatially and temporally adaptive Voronoi cells, that can better represent areas with higher and lower density.

The simulation with higher Mach number $\mathcal{M} = 1$ differs from the others. In particular, the high temperature peak shifts to lower values, suggesting that the phases are more thoroughly mixed due to the higher fluctuation velocities, resulting in more intermediate temperature gas. In general, the temperature PDF of turbulent radiative gas depends not only on Mach number, but also on geometry, as well as the ratio of the mixing to cooling timescales \citep{chen26}.

Figure \ref{fig:phase} also shows the density-temperature phase space diagram at the end of the \texttt{gthmM0.5} simulation (right panel). We see that the cold phase is the densest, while the hot phase favors lower densities. This relationship holds down to temperatures of \(\mathrm{10^{4.3}\,\mathrm{K}}\), beyond which a pile-up occurs towards $\sim 10^4$\,K due to the rapidly declining strength of the cooling function, leading to quasi-stable gas. In addition, we recall our approach to enforced global thermal equilibrium in the simulation. Since the net loss in internal energy is added back in with mass weighting, the densest gas receives a higher portion, counteracting any further cooling.

\subsection{Cold Gas Formation and Evolution}

Turbulence can seed the density fluctuations that enable cold gas to form through the classical mechanism of thermal instability (TI). Slightly over-dense gas cools faster than its surroundings, losing pressure and undergoing compression and runaway cooling, until it reaches a new equilibrium. The classical criterion for TI in a uniform medium with net cooling function \(\Theta = \mathcal{L} - Q\), for thermal equilibrium ($\Theta = 0$), is given by $\left(\partial \Theta / \partial S\right)|_A > 0$ \citep{1965ApJ...142..531F}, where $S$ is the entropy and $A$ is a quantity (pressure or volume) held constant.

For a cooling function $\Lambda(T)$ with local power-law \(\alpha = \mathrm{d} \log \Lambda / \mathrm{d} \log T,\) the above condition reduces to $\alpha < 2$ for isobaric and $\alpha < 0$ for isochoric thermal instability. In the isobaric case for example, a temperature decrease can lower $\Lambda(T)$, but the density increase raises $\mathcal{L}$ through its density dependence. The condition $\alpha < 2$ ensures that the reduction in $\Lambda$ is insufficient to counteract the increase, leading to runaway cooling. The thermal instability timescale at which TI can develop based on the above criterion and a density dependence of the heating rate density \(Q \propto \rho^\alpha\) is \citep{mohapatra23}:
\begin{align}
    t_\mathrm{ti} = \frac{\gamma t_\mathrm{cool}}{2 - \mathrm{d}\log\Lambda/\mathrm{d}\log T - \alpha}.
    \label{eq:thermalinst-tscale}
\end{align}
The cooling timescale \(t_\mathrm{cool}\) is defined as the time gas with a specific internal energy \(e\) needs to loose its energy through radiative cooling given an instantaneous volumetric cooling rate \(\Lambda(T) [\mathrm{erg}\mathrm{s}^{-1}\mathrm{cm}^3]\), thus an ideal gas with an energy loss of \(\mathcal{L} = n_e n_H \Lambda(T)\):
\begin{align}
    t_{\mathrm{cool}} = \frac{\rho e}{\mathcal{L}} = \frac{p}{\left(\gamma - 1\right)n_e n_H \Lambda(T)}.
    \label{eq:coolingtime}
\end{align}
It is not clear if this condition is directly applicable to our turbulent simulations, or if the thermal instabilities are isobaric. At the resolution scale, cooling is isochoric, since the cooling term introduced to the energy conservation equation reduces the internal energy of a cell at a given time step without changing its volume. Subsequently, the pressure of the gas cell must also be updated, such that the cooling is not locally isobaric. It is possible that cool gas moves and exchanges energy to balance out pressure gradients faster than the cooling time.

\begin{figure*}
    \centering
    \includegraphics[width=0.49\textwidth]{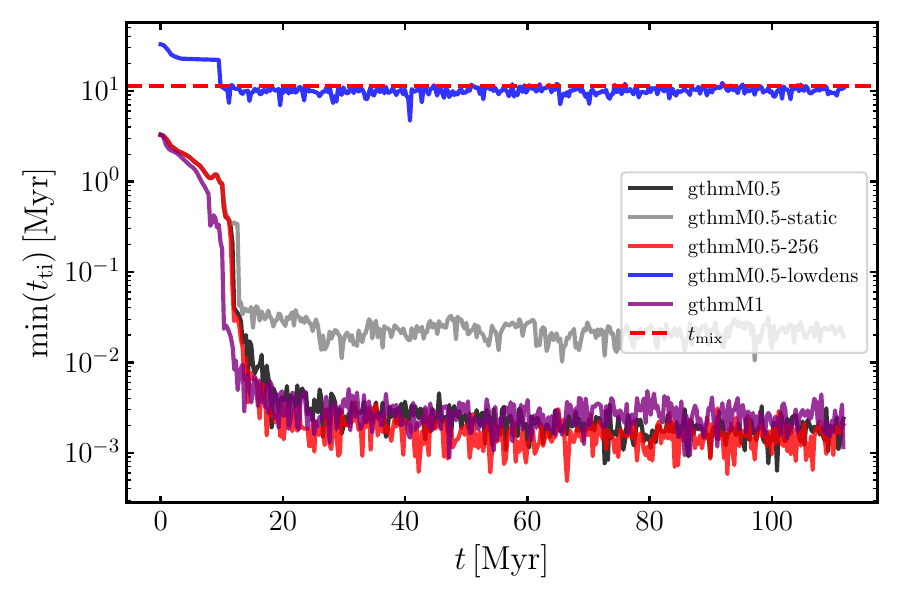}
    \includegraphics[width=0.46\textwidth]{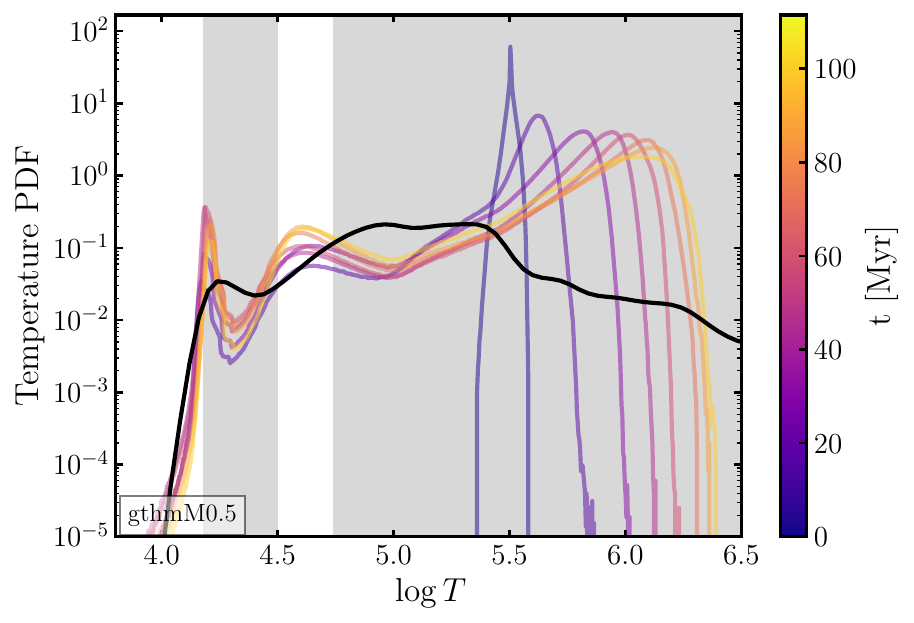}
    \caption{The minimum calculated value of the thermal instability timescale as a function of simulation time (left panel), and the time evolution of the volume-weighted temperature PDF (right panel). In the latter case we focus on the \texttt{gthmM0.5} simulation. The black line represents the temperature dependent cooling function, while the grey shaded regions indicate the temperature ranges in which the slope $\alpha$ (Eqn. \ref{eq:thermalinst-tscale}) satisfies the condition for isobaric thermal instability.}
    \label{fig:ti}
\end{figure*}

Nonetheless, Figure \ref{fig:ti} shows the minimum thermal instability timescale in each simulation as a function of time (left panel).\footnote{The thermal instability timescale is proportional to the cooling time scale of the gas. The proportional constant is of order of unity if the local power-law slope \(\alpha\) is smaller than unity. As \(\alpha\) approaches two, the threshold for isobaric TI, the TI timescale approaches infinity. In the calculation, we therefore set $t_{\rm TI}$ to infinity if \(\alpha\) becomes greater than two.} The minimum fluctuates around a quasi-stationary value, reflecting the gas cell with the highest density at \(T \simeq 10^{4.3}\,\mathrm{K}\), where the combination of cooling rate, its slope and density combines to the minimal value. 

\citet{mohapatra23} suggest that the ratio of thermal instability timescale and turbulent mixing timescale is a criterion for the formation of a multiphase medium. They argue that while turbulence can create density fluctuations, these can also be mixed away if the mixing timescale \(t_\mathrm{mix} = L/V\), given by the characteristic velocity \(V\) and length scale \(L\) of the largest turbulent motions, is short enough. In Figure \ref{fig:ti}, the red dashed line indicates the mixing timescale, calculated as the timescale of turbulent eddies with \(L = L_\mathrm{driv} = 0.5\,\mathrm{kpc}\) and \(V = 42.78\,\mathrm{km}/\mathrm{s}\), which are the turbulent injection scale and expected velocity due to the turbulent driving. This result suggests why the simulation with lower initial density fails to form multiphase gas, since the TI timescale is roughly equal to $t_{\rm mix}$ \citep[e.g.][]{fielding22,abruzzo22,abruzzo24}.

We also find that the minimal thermal instability timescale is roughly an order of magnitude larger for the static mesh simulation. This reflects the previous discussion that Lagrangian methods such as the moving mesh runs are able to resolve gas with higher densities \citep{price10}, and therefore shorter TI timescales. 

All our cooling simulations have an initial temperature distribution that is a delta function at \(T = 10^{5.5}\,\mathrm{K}\). Figure \ref{fig:ti} shows how it develops three distinct peaks over time, reflecting the distinctive shape of the cooling function (right panel). The black line gives the qualitative behavior of the cooling rate as a function of temperature. The grey regions indicate the temperature ranges where the criterion \(\alpha < 2\) is satisfied. There is a large gap centered at \(T \simeq 10^{4.6}\,\mathrm{K}\), where gas is thermally stable based on this criterion. The same behavior occurs at \(\sim 10^{4.2}\,\mathrm{K}\), where the cooling curve drops off quickly, leaving most of the gas at these temperatures. The third peak at high temperatures (\(T > 10^{5.5}\,\mathrm{K}\)) shifts towards even higher temperatures with time. This is likely a consequence of the artificial heating introduced to maintain global thermal equilibrium. The cooling rate at these high temperatures is already low, so the heating overpowers the cooling, leading to a gradual increase in temperature.

As a result of radiative energy loss, the initially hot gas quickly forms a cooler counterpart. Figure \ref{fig:coldmassfrac} shows the mass fraction of cool (\(T < 10^{4.3}\,\mathrm{K}\)) and intermediate temperature (\(T < 10^{5}\,\mathrm{K}\)) gas as a function of time (left panels). These temperature ranges are selected to include gas up to the first and second peak of the temperature distributions, respectively. They trace gas significantly colder than the temperature of the CGM of the virialized halos in rough hydrostatic equilibrium. Impressively, we find that gas with temperature lower than \(10^5\,\mathrm{K}\) makes up roughly 50\% of the total mass by the end of the simulations, in the cases where a multiphase medium develops. There are large fluctuations in the total mass of gas below \(10^{4.3}\,\mathrm{K}\), and it is unclear if a true steady state has been reached in this respect by $10T$.

Physically, over-dense gas has a shorter cooling timescale than its surroundings and can undergo rapid compression, loosing buoyancy support and subsequently falling onto a central galaxy. This process is often labeled `precipitation' or `condensation' and tends to occur if the cooling timescale \(t_\mathrm{cool}\) of the gas is much less than the free fall time \(t_\mathrm{ff}\) \citep{sharma12}. As a result, the production of cool overdensities in turbulent, CGM-like conditions has direct implications for the gas supply responsible for star formation in galaxies.

We also find that there is less cold gas in the simulation with a static mesh. One potential cause is that the moving mesh is able to resolve the clouds in more detail, i.e. with more complex surfaces with correspondingly larger surface area, leading to more efficient mixing of hot and cold gas \citep{agertz07,Fielding2020}. The ability to resolve small-scale individual clouds is a demanding numerical discriminator. In addition to high numerial resolution, we find that the cumulative amount of cold gas below the threshold for the cloud identification is similar between the \texttt{gthmM0.5} and \texttt{gthm0.5-static} simulations. However, the distribution of cloud masses shows that there are fewer clouds of small mass, but more clouds with intermediate mass, in the simulation with a fixed grid. This is likely because the moving mesh is able to form (and advect) clouds that are smaller than one resolution element in the fixed grid simulation with the same density, resulting in lower-mass structures \citep[see also][]{Bauer_2012,cernetic24}.


\section{Discussion} \label{sec_discussion}

\subsection{The morphology of cool gas}

\begin{figure*}
    \centering
    \includegraphics[width=0.98\textwidth]{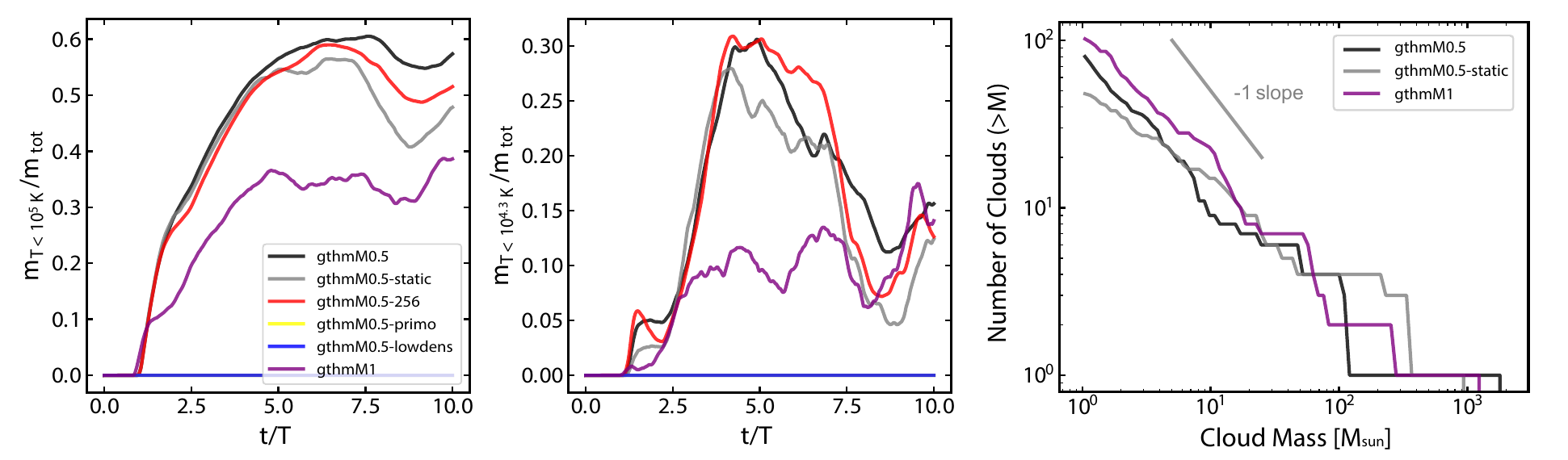}
    \caption{The left two panels show the evolution of the mass fraction of gas at intermediate (\(T < 10^5\,\mathrm{K}\)) and low (\(T < 10^{4.3}\,\mathrm{K}\)) temperature. The right panel shows the cumulative distribution of cloud masses at the end of the simulation. For a given mass \(m\) we calculate the number of cold clouds with mass greater than or equal to \(m\), and find a slope of the cumulative mass function of $\alpha \simeq -1$.}
    \label{fig:coldmassfrac}
\end{figure*}

Beyond global abundance, we move to identify individual clouds of cold gas, allowing us to measure their size and mass distributions. To do so we isolate the peak at the lower end of the temperature distribution, and define gas cells as belonging to a cool cloud if their temperature is below \(10^{4.3}\,\mathrm{K}\). Although this could artificially separate structures connected through intermediate temperature gas, we find that the relative sharpness of interfaces negates this issue. Clouds are then identified as the collection of connected groups of Voronoi cells with temperature below the threshold  \citep[following][]{Nelson-coldgas,ramesh2023}. 

Figure \ref{fig:coldmassfrac} shows the resulting cumulative mass distribution of individual cool clouds (right panel). The overall trend shows that number of clouds with a mass greater than \(m\) roughly scales with \(m^{-1}\), meaning that most of the clouds have low masses and only few have a mass comparable to the total mass (\(= \rho_0 V = 1.18\times10^4\,\mathrm{M}_\odot\)) contained in the simulation. Since the cumulative power-law slope is always one less than the corresponding non-cumulative distribution, this corresponds to a slope of $dN/dm \propto m^{-2}$ for the differential cloud mass function. Comparing the $\mathcal{M} = 1$ (purple) and $\mathcal{M} = 0.5$ (black) cases, we find hints that the mass function slope may steepen with increasing turbulent Mach number, and/or in the transition to the supersonic regime. In general, the shape of the mass (and size) distributions of cool circumgalactic clouds is an ongoing topic of study.

The slope we identify herein is in quantitative agreement with the cloud mass function found in the TNG50 cosmological simulation \citep{nelson20}, as well as in the GIBLE CGM refinement simulations of significantly higher resolution \citep{ramesh2023}. Likewise, \citet{tan24} identify a remarkably similar cumulative volume scaling $dN (>V) / dV \propto V^{-1}$, while concluding that fragmentation of the initial ISM of a disk galaxy during supernovae-driven superbubble breakout determines the distribution of cloud sizes. As our simulations contain no ISM, no galaxies, and no supernovae, the commonality of these results suggests a more universal origin related to turbulent flows.

Indeed, power-law distributions with slopes of $-2$ are a common feature in many astrophysical phenomena, particularly in regimes of gravitational collapse and structure formation across scales, from dark matter halos to stars \citep{guszejnov18}. Such a scaling relation indicates that there is a constant contribution (i.e. mass) per logarithmic decade. The idealized turbulence simulations of \citet{gronke22} likewise identify a $dN/dm \propto m^{-2}$ mass distribution for small-scale cool clouds, and identify cloud-cloud interactions and coagulation as important processes in establishing this slope. The MHD turbulence simulations of \citet{fielding23} find the same scale-free mass distribution, albeit with a tendency for a single large cloud to dominate in terms of total mass.

It is clear that the morphological details of multi-phase gas result from the non-trivial interplay of multiple physical processes. It is also reasonable that, despite suggestions towards universality for cool CGM clouds, quantitative descriptions will depend on the details of the physics at play.

\subsection{Potential Caveats and Future Directions}

Our simulations include turbulent driving and radiative cooling of a plasma in collisional ionization equilibrium (CIE) at solar metallicity. We discuss the realism, potentially missing physics, and applicability of these simulations to the CGM as a whole.

Our fiducial setup adopts a mean density that is ten times higher than the median density value obtained from our analysis of the CGM of TNG50 Milky-Way like galaxies. We did so to decrease the cooling timescale of the gas, facilitating the rapid formation of a multiphase medium. Therefore, the simulations could represent a less common, higher density region of the CGM. Alternatively, we could also promote the formation of multiphase gas by starting with a broader initial distribution of densities, at the same mean density. This would initialize regions with different cooling timescales, and is motivated by the fact that the CGM contains hydrodynamical shocks and even cool gas that has not formed in-situ but has been ejected in outflows \citep{Nelson-coldgas,almeida26}, accreted onto the halo from the intergalactic medium \citep{dekel09,nelson13}, or been stripped from incoming satellite galaxies \citep{ramesh23,decataldo24}. 

In addition, we invoke a cooling function based on CIE at solar metallicity. CIE itself is a dubious assumption, particularly given the importance of photoionization from the metagalactic UVB \citep{byrohl23,taira25}, in addition to the potential impact of local radiative sources \citep{oppenheimer18,zhu24,cadiou25}. More realistic non-equilibrium ionization state solvers for diffuse gas may be critical for accurate cooling physics \citep{katz22}. In terms of metallicity, this is on the high side for gaseous halos at the Milky Way mass scale \citep{Prochaska_2017}, but reasonable for outflow-enriched disk-CGM interface regions \citep{Tumlinson_2017}. Metallicity inhomogeneities are also present, and could lead to regions of the CGM where metal cooling is more versus less important. Although we did not measure gas-phase metallicity in our TNG50 analyses, this is a straight-forward extension. 

We also highlight our invocation of global thermal equilibrium in the simulations. The CGM of more massive halos is supported in hydrostatic equilibrium. This does not imply thermal equilibrium, but there are heating mechanism that counteract the formation of a single phase cooling flow in the CGM \citep{fauchergiguere2023}. We chose to focus on the formation of a multiphase medium with local thermal instabilities \citep[e.g.][]{sharma10}, but the need for such a trick suggests that key heating processes, such as galactic feedback or large-scale accretion, are absent.

\subsection{Future Directions}

First, our current simulations do not include magnetic fields. These can be important for the dynamics of a turbulent and radiative cooling plasma \citep{lehle26}. For instance, magnetic fields suppress mixing in turbulent mixing layers between hot and cold phases \citep{Ji__2019}. Using TNG50, \citet{Nelson-coldgas} show that cold clouds in the CGM are severely thermally underpressurized but have high magnetic pressure, meaning that magnetic fields can dominate the dynamics. There have also been extensions to account for the relationship between Mach number and density fluctuations in a turbulent plasma including magnetic fields. \citet{molina2012} find a density contrast dispersion Mach number relation similar to Eqn. \ref{eq:basic_densmach} containing the plasma \(\beta\): \(\sigma^2_s = \ln\left(1 + b^2\mathcal{M}^2\beta/(\beta + 1)\right)\). Turbulence in magnetohydrodynamics has additional complications, giving rise to anisotropy in the subsonic regime and strongly affecting the turbulent energy cascade \citep{beattie2024}. Magnetic fields should clearly be included in simulations of the multiphase CGM. We have measured $\beta$ values from TNG50 in preparation for extending our current simulations with the ideal MHD scheme available in AREPO, but leave this exploration for future work.

The gas in the CGM is also stratified, with significant radial structure. Turbulence in a stratified medium is different, since kinetic energy transform into potential energy, and the effect of buoyancy plays a key role in the dynamics. \citet{Mohapatra_2020} extend the density contrast variance Mach number relation to include the impact of stratification. Similarly, as a first approximation, our local turbulent boxes could be extended with a uniform gravitational field.

Finally, the nature of our turbulent energy injection scheme has several simplifications. We assume constant and isotropic turbulent energy injection. However, turbulence in the CGM is undoubtedly sourced from different (and multiple) physical mechanisms, with correspondingly different scales. It is not necessarily stationary, and the physics of the onset and decay of turbulence could further influence the morphology of the CGM. Fully capturing this physics requires self-consistent and simultaneous simulation of the cosmological context and the central galaxy, at a minimum. The ultra-high resolution CGM refinement techniques offer a potential path forward, as to the recent `cosmological windtunnel' setups that reach parsec and solar mass scales in fully cosmological setups \citep{ramesh26}. Such approaches offer the possibility of more fully resolving the turbulent cascade and the formation and evolution of small-scale gas structure in the circumgalactic medium of galaxies.


\section{Summary and Conclusions} \label{sec_conclusions}

In this paper we study turbulent multiphase media in the circumgalactic medium (CGM) of galaxies. We first analyze the gas properties and turbulent kinematics in the CGM of Milky Way-like galaxies from the TNG50 cosmological magnetohydrodynamical simulation. These sample the relevant parameter space, and provide us with realistic initial conditions. We then implement a turbulent driving routine and metal-line radiative cooling physics into the public version of the \texttt{Arepo} code, and run a suite of idealized, $1$\,kpc$^3$ turbulent boxes of CGM-like media. Our main results are:

\begin{itemize}
    \item To analyze the turbulent kinematics of TNG50 halos, we apply a multi-scale filtering approach to separate turbulent from bulk velocities, plus a Helmholtz decomposition to isolate the compressive component. We find that the CGM of Milky Way-like galaxies at $z=0$ contains predominantly subsonic flows (Mach number $\mathcal{M} \sim 0.1 - 0.7$ and $\bar{\mathcal{M}} \simeq 0.3$) that are roughly balanced between compressive and solenoidal components with $b = \mathcal{M}_{\rm comp}/\mathcal{M}_{\rm tot} \sim 0.3 - 0.6$.
    \item We implement a method to drive statistically stationary and homogeneous turbulence in idealized simulations. The variance and autocorrelation of the forcing field is set by an Ornstein-Uhlenbeck process at low spatial frequency (large scales).
    \item We verify the driven turbulence setup with a series of isothermal simulations, varying the turbulent Mach number and compressiveness of the forcing. We find rapid stabilization of density and velocity statistics after $\sim 2-3$ autocorrelation times. Turbulent motions below the injection scale develop self consistently, and the (small) inertial range has a velocity power spectrum consistent with expectations from incompressible turbulence theory.
    \item The isothermal density PDF is broader at larger Mach number, and for more compressive driving, in agreement with previous work. The compressive parameter $b$ decreases with increasing $\zeta$, the spectral weight parameter, as expected from previous work on supersonic isothermal turbulence. However, for subsonic turbulence, we find that the forcing is less effective at creating compressive motion.
    \item We then implement a treatment for metal-line (CIE) radiative cooling, and run a suite of turbulent box simulations with cooling, maintaining global thermal equilibrium. When the thermal instability timescale $t_{\rm TI}$ is sufficiently short, we find that initially uniform and warm ($T = 10^{5.5}$\,K) gas spontaneously forms overdense and cool filamentary and cloud-like structures. Depending on the simulation, the cool phase can dominate, saturating at mass fractions $\gtrsim 50$\%. When $t_{\rm TI}$ is instead comparable to the mixing timescale $t_{\rm mix}$ at the turbulent injection scale, cool gas fails to form.
    \item We identify individual cold clouds using the Voronoi tessellation. Comparing moving to static mesh simulations, we find a clear preference for smaller, denser, and more geometrically complex cold structures in the moving mesh case. The cumulative cool cloud mass function is consistent with a power-law slope of $dN(>M)/dM \propto M^{-1}$ i.e. $dN/dM \propto M^{-2}$. We tentatively identify a steepening dependence of the mass function slope with increasing turbulent Mach number.
\end{itemize}

Our simulations set the stage for resolving the small-scale physics of turbulence, radiative cooling, and mixing in cosmologically realistic CGM-like conditions. In the future, we can more fully identify the parameter space where local thermal instabilities lead to the formation of a multiphase medium. This will allow us to holistically assess the observational implications of turbulent multi-phase gas in the gaseous halos surrounding galaxies, in terms of both absorption and emission signatures. In addition, our results can serve as a foundation for developing sub-grid models for future galaxy formation simulations, where the impact of turbulence and small-scale cold gas structures are unresolved \citep{butsky25,smith24,das24}.


\section*{Data Availability}

The IllustrisTNG simulations, including TNG50, are publicly available and accessible at \url{www.tng-project.org/data} \citep{nelson19a}. Data related to this publication is available on request from the authors. The modified version of \texttt{Arepo} with our turbulent driving routine is publicly available.\footnote{\url{www.github.com/nelson-group/jbiba-arepo}} The turbulence analysis routines and Helmholtz decomposition method are available as part of the \texttt{temet} package (\textcolor{blue}{Nelson in prep}).\footnote{\url{www.github.com/dnelson/temet}}

\section*{Acknowledgements}

The authors thank Chris Byrohl, Annalisa Pillepich, and Bipradeep Saha for helpful discussions. DN acknowledges funding from the Deutsche Forschungsgemeinschaft (DFG) through an Emmy Noether Research Group (grant number NE 2441/1-1). This work is supported by the Deutsche Forschungsgemeinschaft (DFG, German Research Foundation) under Germany's Excellence Strategy EXC 2181/1 - 390900948 (the Heidelberg STRUCTURES Excellence Cluster). The simulations have made use of the Vera cluster of the Max Planck Institute for Astronomy (MPIA), operated by the Max Planck Computational Data Facility (MPCDF).

\bibliographystyle{aa}
\bibliography{refs}


\appendix

\section{Helmholtz Decomposition}

\begin{figure*}
    \centering
    \includegraphics[width=0.9\textwidth]{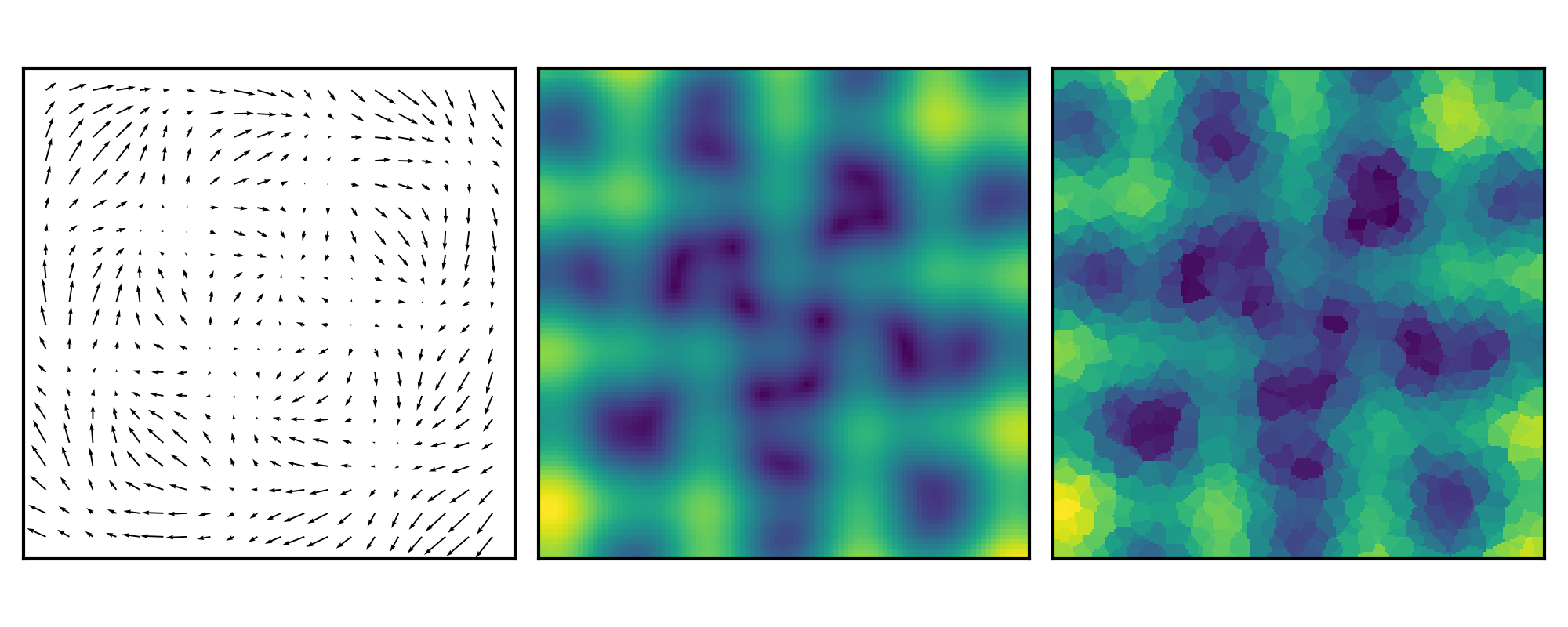}
    \caption{An example vector field for the demonstration of the Helmholtz decomposition. Left: Quiver plot of a 2D slice of the vector field. Center: Magnitude of the vector field of the same slice. Right: The same vector field but defined on a Voronoi mesh.}
    \label{fig:vector_field_toy}
\end{figure*}

To decompose a vector field in its purely solenoidal and compressive components, one can apply the projection tensors $\mathcal{P}^{\parallel}_{ij}(\bm{k}) := k_i k_j/k^2$ and $\mathcal{P}^{\perp}_{ij}(\bm{k}) := \delta_{ij} - k_i k_j/k^2$ in Fourier space. Let $\bm{v}$ be a vector field and
\begin{align}
\hat{\bm{v}}(\bm{k}) = \int \bm{v}(\bm{x})e^{-i \bm{k}\cdot \bm{x}}\mathrm{d}^3x
\end{align}
its Fourier transform. If we define
\begin{align}
\hat{\bm{v}}^\parallel(\bm{k}) &:= \mathcal{P}^{\parallel}(\bm{k})\hat{\bm{v}}(\bm{k}) = \hat{\bm{v}}(\bm{k}) - \hat{\bm{v}}(\bm{k}) \cdot \bm{k} \frac{\bm{k}}{k^2} \\
\hat{\bm{v}}^\perp(\bm{k}) &:= \mathcal{P}^{\perp}(\bm{k})\hat{\bm{v}}(\bm{k}) = \hat{\bm{v}}(\bm{k}) \cdot \bm{k} \frac{\bm{k}}{k^2}, 
\end{align}
we have:
\begin{align}
\hat{\bm{v}}(\bm{k}) = \hat{\bm{v}}^\parallel(\bm{k}) + \hat{\bm{v}}^\perp(\bm{k})\nonumber\\
\bm{k} \cdot \hat{\bm{v}}^\parallel(\bm{k}) = 0\\
\bm{k} \times \hat{\bm{v}}^\perp(\bm{k}) = 0\nonumber
\end{align}
As a result,
\begin{align}
\bm{v}(\bm{x}) &= \int \hat{\bm{v}}(\bm{k})e^{i\bm{k}\cdot \bm{x}} \frac{\mathrm{d}^3k}{\left(2\pi\right)^3} \\&= \int \left(\hat{\bm{v}}^\parallel(\bm{k}) + \hat{\bm{v}}^\perp(\bm{k})\right)e^{i\bm{k}\cdot \bm{x}} \frac{\mathrm{d}^3k}{\left(2\pi\right)^3} \nonumber\\ &= \int \hat{\bm{v}}^\parallel(\bm{k})e^{i\bm{k}\cdot \bm{x}} \frac{\mathrm{d}^3k}{\left(2\pi\right)^3} + \int \hat{\bm{v}}^\perp(\bm{k})e^{i\bm{k}\cdot \bm{x}} \frac{\mathrm{d}^3k}{\left(2\pi\right)^3} \nonumber \\&=: \bm{v}^\parallel(\bm{x}) + \bm{v}^\perp(\bm{x}) \nonumber\\
\bm{\nabla} \cdot \bm{v}^\parallel(\bm{x}) &= \int \bm{\nabla} \cdot \left(\hat{\bm{v}}^\parallel(\bm{k})e^{i\bm{k}\cdot \bm{x}}\right)\frac{\mathrm{d}^3k}{\left(2\pi\right)^3} \\&= \int i \bm{k} \cdot \hat{\bm{v}}^\parallel(\bm{k})e^{i\bm{k}\cdot \bm{x}}\frac{\mathrm{d}^3k}{\left(2\pi\right)^3} = 0 \nonumber\\
\bm{\nabla} \times \bm{v}^\perp(\bm{x}) &= \int \bm{\nabla} \times \left(\hat{\bm{v}}^\perp(\bm{k})e^{i\bm{k}\cdot \bm{x}}\right)\frac{\mathrm{d}^3k}{\left(2\pi\right)^3} \\&= \int i \bm{k} \times \hat{\bm{v}}^\perp(\bm{k})e^{i\bm{k}\cdot \bm{x}}\frac{\mathrm{d}^3k}{\left(2\pi\right)^3} = 0.\nonumber
\end{align}

To perform a Helmholtz decomposition on unstructured data we use a FFT to transform the vector field to a grid in Fourier space and then perform a decomposition with the projection operators $\mathcal{P}^{\parallel}_{ij}(\bm{k}) := k_i k_j/k^2$ and $\mathcal{P}^{\perp}_{ij}(\bm{k}) := \delta_{ij} - k_i k_j/k^2$ \citep[following][]{Bauer_2012,2021MNRAS.504..510V}. To do so we must sample the values defined on the Voronoi grid onto a cartesian grid, similarly as for our visualization method.

To test this approach we define a toy scalar and vector potential \(\phi\) and \(\bm{A}\) and derive a vector field \(\bm{v} = \nabla\phi + \nabla\times\bm{A}\) with finite differencing, as shown in Figure \ref{fig:vector_field_toy}. In this case we know the correct decomposition. We then replicate our typical scenario by defining a random set of points to be Voronoi mesh generators and pick the value for the corresponding Voronoi cell based on the closest point of the vector field. We then perform a Helmholtz decomposition of this vector field, as if it were the output of an \texttt{Arepo} simulation.

\begin{figure*}
    \centering
    \includegraphics[width=0.9\textwidth]{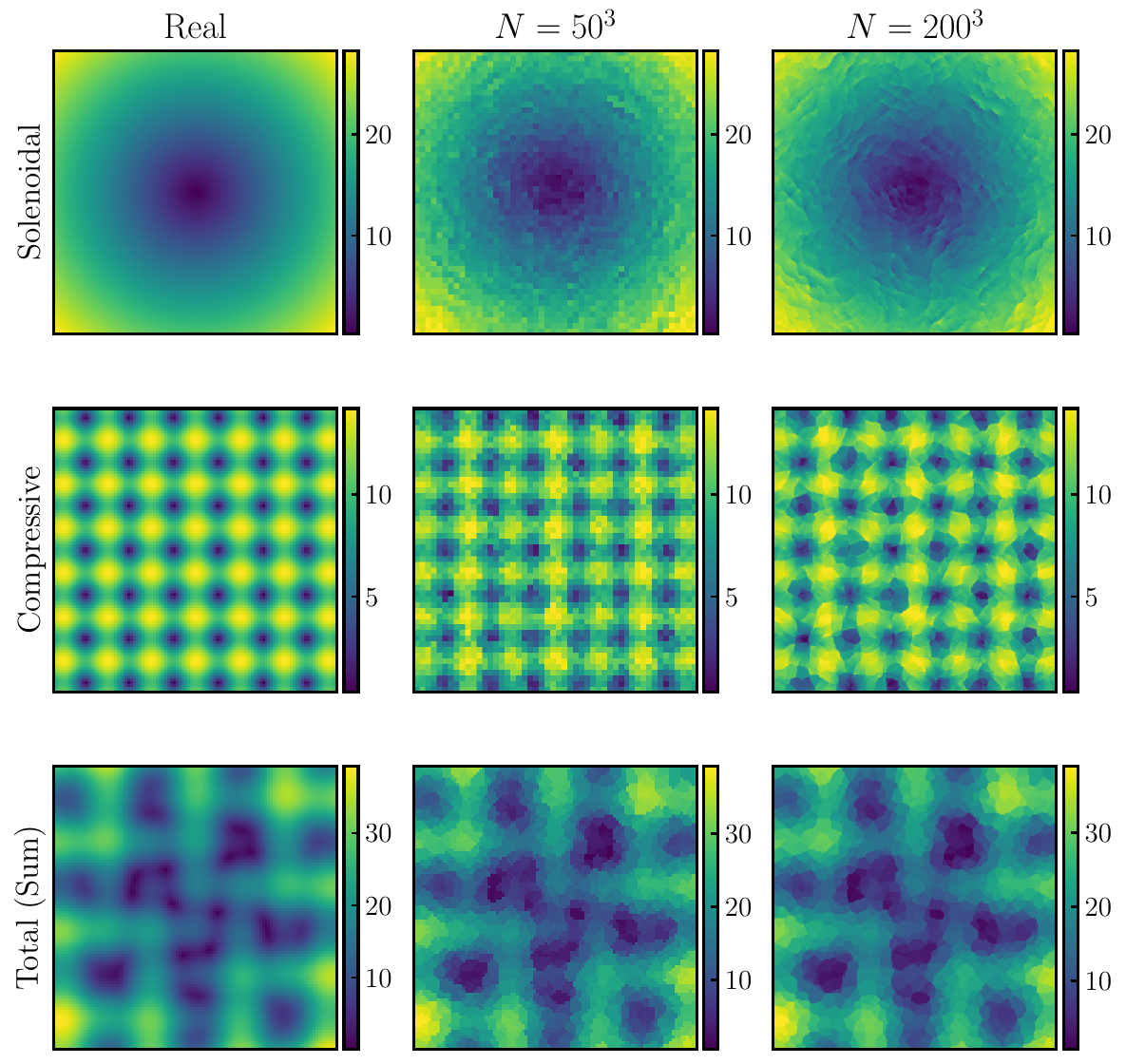}
    \caption{Comparison of the decomposition method. The test vector field is remapped to a cartesian grid with two different resolutions. The rows are from top to bottom: the solenoidal part of the vector field, the compressive part and the total vector field (the sum of solenoidal and compressive). The left column shows the "real" vector field obtained through finite differencing of the given scalar and vector potentials. The second and third columns show a remapping and decomposition with a fixed grid of \(50^3\) and \(200^3\) points.}
\end{figure*}

We verify that the impact of the sampling grid resolution on measurements of the RMS of the vector field are negligible and converge rapidly.

\section{Power Spectra Measurement}

We measure the power spectrum of a quantity \(\omega(\bm{x})\) following \citet{Bauer_2012}. The power spectrum is defined as the Fourier transform of the two-two point correlation function \(C(\bm{l}) = \langle\omega(\bm{x})\omega(\bm{x}+\bm{l})\rangle_{\bm{x}}\). We use this convention for the Fourier transform of a periodic quantity with period \(L\) in all spacial directions:

\begin{align}
    \hat{\omega}(\bm{k}) = \frac{1}{(2\pi)^3}\int_V\omega(\bm{x})e^{-i\bm{k}\bm{x}}\mathrm{d}^3x.
\end{align}

The power spectrum is then defined as the Fourier transform of the two-point correlation function:

\begin{align}
    P(\bm{k}) = \frac{1}{(2\pi)^3}\int_V C(\bm{l})e^{-i\bm{k}\bm{l}}\mathrm{d}^3l = \left(\frac{2\pi}{L}\right)^3 \left|\hat{\omega}(\bm{k})\right|^2.
\end{align}

If the statistical distribution of \(\hat{w}(\bm{k})\) is isotropic, we can calculate the 1D power spectrum by angular averaging over radial shells in \(\bm{k}\)-space:

\begin{align}
    P(k) = 4\pi k^2\langle P(\bm{k})\rangle.
\end{align}

In practice one can do this by defining logarithmic bins in \(k\) to calculate the mean of all the values falling in that bin. The power spectrum is finally normalized such that the integral over the 1D power spectrum gives the variance of the field used to measure the spectrum according to Parseval's identity:

\begin{align}
    \int P(k)\mathrm{d}k = \frac{1}{N^3} \sum_{i, j, k = 0}^{N-1}\left|\omega_{ijk}\right|^2.
\end{align}

We test this method in the following scenario. A vector field is set up in Fourier space on a fixed grid with amplitudes based on the wave number \(k\), such that the one dimensional power spectrum has a power-law slope of \(-5/3\). As described in the sections before, we can simulate that this vector field is given on a Voronoi grid by selecting a number of randomly distributed points to be the mesh generators, where the value of the vector field is given by its value of the closest point in the original fixed grid. 

\begin{figure*}
    \centering
    \includegraphics[width=0.9\textwidth]{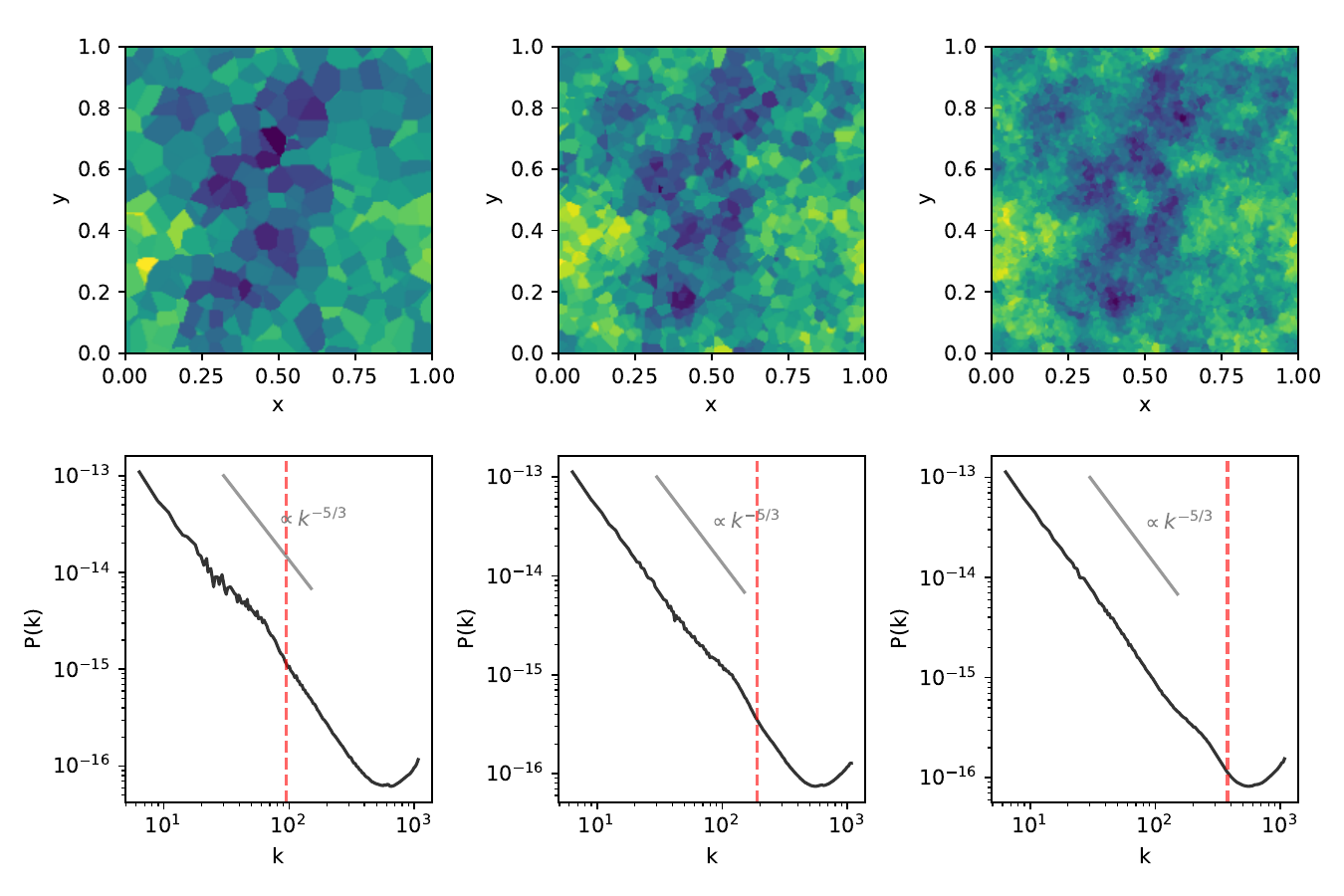}
    \caption{Comparison of our power spectrum measurement method for different Voronoi grid resolutions. From left to right: \(15^3\), \(30^3\) and \(60^3\) random points are used for generating the Voronoi mesh. The first row shows a slice of the magnitude of the vector field, while the second row shows the calculated one dimensional power spectrum. The dashed line in the power spectra marks the scale of the average separation of the Voronoi cells. The slope is in rough agreement with the original for the low wave numbers even in if the number of mesh generating points is as low as \(15^3\).}
    \label{fig:powerspec-vorrescomp}
\end{figure*}

\subsubsection{Resolution Convergence with Cooling}

\begin{figure*}
    \centering
    \includegraphics[width=0.8\textwidth]{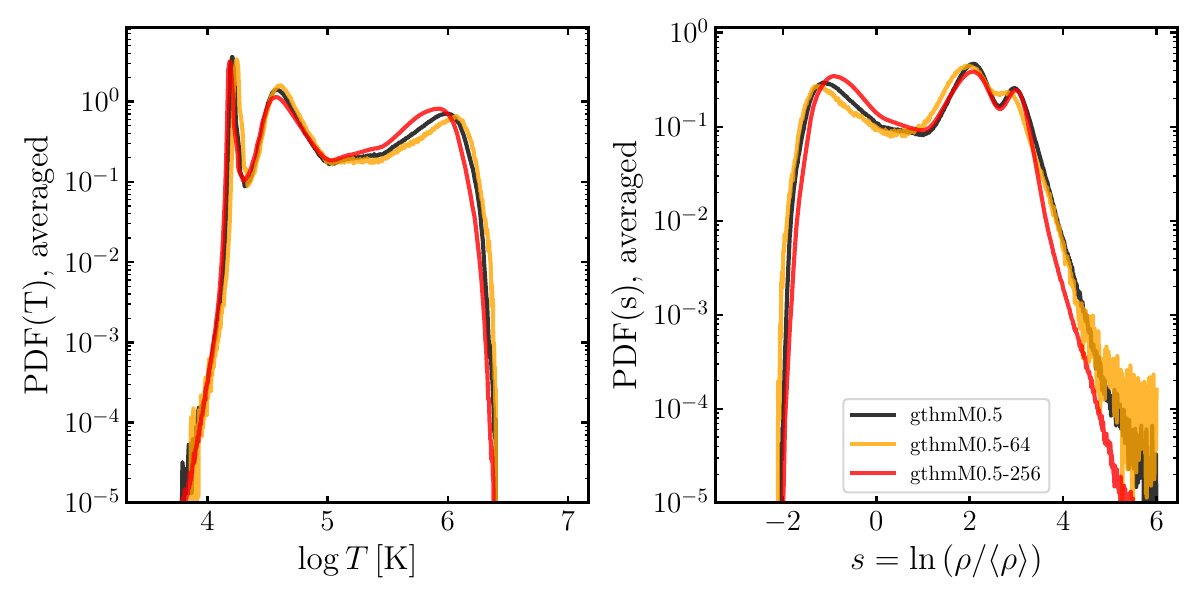}
    \caption{Volume weighted density contrast and temperature PDFs of three simulations with the same setup and initial conditions but at three different resolutions.}
    \label{fig:resolution-denstemppdf}
\end{figure*}

We examine the numerical convergence of our idealized turbulent box simulations where it may be challenging: with the inclusion of radiative cooling. To do so, we compare the \texttt{gthmM0.5} and \texttt{gthmM0.5-256} simulations, together with a \(64^3\) resolution run with the same setup and initial conditions. Figure \ref{fig:resolution-denstemppdf} shows the temperature and density PDFs averaged over the final 10 snapshots of each.

The overall shape is impressively consistent across the three levels of resolution. In addition, there are some second order resolution effects. In particular, the peak at higher temperatures in the PDF shifts slightly towards lower temperatures, while the low density peak in the density PDF moves towards higher densities and the high density tail drops off more quickly at higher resolution. 

\subsubsection{Magnetic Field Decomposition}

\begin{figure*}
    \centering
    \includegraphics[width=\textwidth]{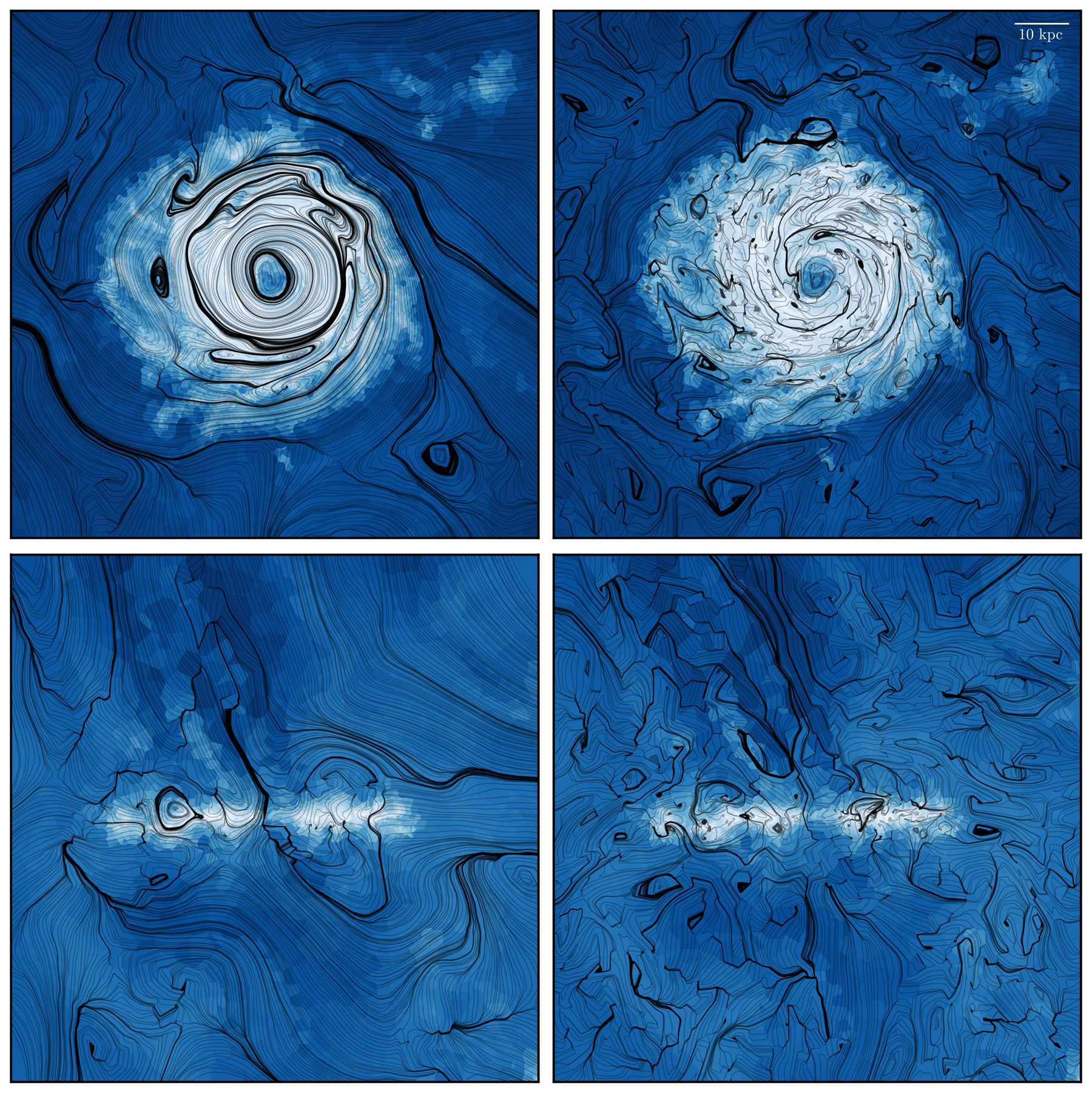}
    \caption{The turbulence decomposition applied to the magnetic fields of a random galaxy in the MWM31 catalog. The illustration shows a slice through in the galactic plane and a slice perpendicular to it. The color represents gas density from dark blue for lower density to white for higher densities. The black lines are an illustration of the magnetic field lines. They are integral curves to the 2d slice of the bulk (left) and turbulent (right) magnetic field.}
    \label{fig:magnetic-turbdecomp}
\end{figure*}

The multi-scale filtering method can be applied to any other 3d vector quantity that has bulk and turbulent component. Thus, one can also use the this method to decompose the magnetic field of the galaxies. Fig. (\ref{fig:magnetic-turbdecomp}) shows a visualization of decomposed magnetic field lines of a galaxy in the MWM31 catalog. The black lines depict the magnetic field lines, traced by following the local direction of the magnetic field. 

\end{document}